\documentclass[superscriptaddress,twocolumn,showpacs,preprintnumbers,amsmath,amssymb,pra,10pt,aps]{revtex4-1}
\usepackage{graphicx}
\usepackage{dcolumn}
\usepackage{bm}
\usepackage{subfigure}
\usepackage{multirow}
\usepackage{soul}
\usepackage{float}
\usepackage[normalem]{ulem}
\usepackage[usenames,dvipsnames]{color}

\def\QE{\textsc{Quantum ESPRESSO}\,}
\definecolor{tangerine}{rgb}{0.944,0.522,0}

\newcommand{\editor}[2]{%
  \expandafter\newcommand\csname #1note\endcsname[1]{%
    \textcolor{#2}{(\textbf{#1:} ##1)}}%
  \expandafter\newcommand\csname #1\endcsname[1]{%
    \textcolor{#2}{##1}}%
  \expandafter\newcommand\csname #1cancel\endcsname[1]{%
    \textcolor{#2}{\sout{##1}}}%
  \expandafter\newcommand\csname #1change\endcsname[2]{%
    \textcolor{#2}{\sout{##1} ##2}}%
  \newenvironment{#1text}{\color{#2}}{\color{black}}
}

\begin{document}

\title {Spin dynamics from \\ time-dependent density functional perturbation theory}
\author{Tommaso Gorni}
\affiliation{
  Scuola Internazionale Superiore di Studi Avanzati (SISSA), Via Bonomea 265, 34136 Trieste, Italy
}
\affiliation{
  Present address: Institut de Min\'eralogie, de Physique des Mat\'eriaux et de Cosmochimie (IMPMC), Sorbonne Universit\'e, CNRS, IRD, MNHN, 4 place Jussieu 75005 Paris, France
}

\author{Iurii Timrov}
\affiliation{
  Scuola Internazionale Superiore di Studi Avanzati (SISSA), Via Bonomea 265, 34136 Trieste, Italy
}
\affiliation{
  Present address: Theory and Simulation of Materials (THEOS), and National Centre for Computational Design and Discovery of Novel Materials (MARVEL), \'Ecole Polytechnique F\'ed\'erale de Lausanne, 1015 Lausanne, Switzerland
}

\author{Stefano Baroni}
\affiliation{
  Scuola Internazionale Superiore di Studi Avanzati (SISSA), Via Bonomea 265, 34136 Trieste, Italy
}
\affiliation{
  CNR-IOM DEMOCRITOS Simulation Center, 34136 Trieste, Italy
}

\date{\today}

\begin{abstract}

We present a new method to model spin-wave excitations in magnetic solids, based on the Liouville-Lanczos approach to time-dependent density functional perturbation theory. This method avoids computationally expensive sums over empty states and naturally deals with the \linebreak coupling between spin and charge fluctuations, without ever explicitly computing charge-density \linebreak susceptibilities. Spin-wave excitations are obtained with one Lanczos chain per magnon wave-number and polarization, avoiding the solution of the linear-response problem for every individual value of frequency, as other state-of-the-art approaches do. Our method is validated by computing magnon dispersions in bulk Fe and Ni, resulting in agreement with previous theoretical studies in both cases, and with experiment in the case of Fe. The disagreement in the case of Ni is also comparable with that of previous computations.

\end{abstract}


\maketitle

\section{Introduction}
\label{sec:intro}

Spin dynamics in magnetic systems is at the core of many interesting phenomena and technologies~\cite{Moriya:1985, Zakeri:2014}. Its thorough characterization has become possible in the last 50 years by the development and continuous refinement of magnetic spectroscopies, most notably  inelastic neutron scattering spectroscopy (INSS) for bulk materials~\cite{Mook:1973}, spin-polarized electron energy loss spectroscopy (SPEELS) and inelastic scanning tunneling spectroscopy for thin films~\cite{Qin:2015, Hirjibehedin:2006}. These spectroscopies allow probing collective and single-particle magnetic excitations due to spin-density fluctuations, namely, spin-wave (magnons) and Stoner (spin-flip) excitations, respectively. At long wave-lengths magnons have lower energies and are long lived. At smaller wavelengths the energies of these two excitation channels become comparable, and magnon lines broaden due to the coupling with spin-flip electron-hole pairs, a process usually referred to as Landau damping~\cite{Landau:1946, Fetter:1971}.

From a theoretical point of view, model Hamiltonians are often used to describe magnetic excitations, possibly in conjunction with \emph{ab initio} results to fit the parameters appearing therein~\cite{Costa:2010, Bergman:2010, Zakeri:2012, Zakeri:2017}. While such models can be derived systematically from an adiabatic decoupling of the spin degrees of freedom from charge fluctuations \cite{Niu:1998,Gebauer:2000} within constrained density functional theory, a fully \emph{ab~initio} treatment of spin-wave dynamics requires the computation of the dynamical spin susceptibility from either
time-dependent density functional theory (TDDFT)~\cite{Savrasov:1998, Lounis:2011, Buczek:2011b, Rousseau:2012, dosSantosDias:2015, Wysocki:2017, Cao:2018} or many-body perturbation theory (MBPT)~\cite{Aryasetiawan:1999, Karlsson:2000, Kotani:2008, Sasioglu:2010, Muller:2016}. Both these methods treat charge and spin fluctuations on an equal footing in a self-consistent manner and they are formally exact, though in practice they rely on different approximations and have different computational requirements. TDDFT is numerically way less demanding than MBPT, particularly when adopting the adiabatic local spin density approximation (ALSDA), which results in a good compromise between computational cost and accuracy~\cite{Runge:1984, Marques:2012} and has in fact been widely adopted for modelling magnetic excitations.
Previous attempts to compute magnon dispersion relations from linear-response TDDFT were based on either the solution of the time-dependent Sternheimer equation \cite{Savrasov:1998, Cao:2018} or of the Dyson equation for the spin susceptibility, starting from the independent-electron spin and charge susceptibilities \cite{Lounis:2011, Buczek:2011b, Rousseau:2012, dosSantosDias:2015, Wysocki:2017}.
In all these approaches the linear-response problem must be solved for every individual value of
the excitation frequency, which is one of the main computational bottlenecks to be
addressed and overcome in this work.

In the present study we introduce a generalization of the Liouville-Lanczos approach to time-dependent density functional perturbation theory (TDDFpT)~\cite{Baroni:2012, Rocca:2008, Timrov:2013}, which allows us to treat the dynamical spin-fluctuation response of magnetic systems in a fully non-collinear framework, and thus model their spin-wave excitation spectra entirely from first principles. Using techniques borrowed from static density functional perturbation theory (DFpT)~\cite{Baroni:1987, Baroni:2001} and similarly to the method of Ref.~\cite{Cao:2018}, our method avoids computing any independent-particle susceptibilities, and thus does not require computationally expensive and slowly converging sums over empty states. At variance with previous studies, our method also avoids repeated linear-response calculations for each individual excitation frequency, by using a recursive Lanczos algorithm to solve the quantum Liouville equation, independently of the frequency. The actual spectrum is then computed upon completion of the compute-intensive Lanczos recursion, in an inexpensive post-processing step for any desired frequency. This allows us to obtain the full spectrum of magnetic excitations (both magnons and Stoner excitations) in any wide frequency range with just one Lanczos chain per excitation wave-number and polarization. 

This paper is organized as follows. In Sec.~\ref{sec:theory} we describe the Liouville-Lanczos approach within TDDFpT for magnetic excitations, in Sec.~\ref{sec:validation} we present a validation of our approach on the prototypical systems bulk Fe and Ni, and in Sec.~\ref{sec:conclusions} we give our conclusions.
Appendix~\ref{sec:general_pert} contains a discussion about a generic perturbing potential which consists of the scalar and magnetic parts, while in the main text only the external magnetic field is considered explicitly.
Lastly, the main text contains a general formulation of the magnetic Liouville-Lanczos approach, while in Appendix~\ref{sec:metals} we give the details necessary to implement it for metals.

\section{Theory and algorithms}
\label{sec:theory}

In INSS experiments a neutron beam with wave-vector $\mathbf{k}_i$ and energy $E_i$ impinges
on the target sample. Due to inelastic scattering, the outgoing neutron will be characterized
by the wave-vector $\mathbf{k}_f = \mathbf{k}_i - \mathbf{q}$ and energy
$E_f = E_i - \hbar \omega$, where $\hbar\mathbf{q}$ and $\hbar \omega$ are the momentum and energy
transferred to the sample, respectively. In the first Born approximation~\cite{Halpern:1938aa,Blume:1963aa}, the double-differential
cross section corresponding to magnetic excitations of electrons can be written in the compact form as:
\begin{equation}
\frac{d^2 \sigma}{d\Omega d\omega} =
\frac{\hbar}{\pi}\left(\frac{g_n e}{2\hbar}\right)^2\frac{k_f}{k_i} \, S(\mathbf{q},\mathbf{\omega}) \,,
\label{eq:cross_section}
\end{equation}
where
\begin{equation}
S(\mathbf{q},\mathbf{\omega}) = -{\rm Im}\,{\rm Tr} \bigg[ {\boldsymbol P}^{\perp}(\mathbf{q}) \,
{\boldsymbol \chi}(\mathbf{q},\mathbf{q}; \omega) \bigg] \,.
\label{eq:S_def}
\end{equation}
Here, $-e$ and $g_n \approx 3.826$ are the electron charge and the neutron $g$-factor, respectively,
${\boldsymbol P}^{\perp}(\mathbf{q})$ is the $3 \times 3$ matrix, $P_{\alpha\beta}^{\perp}(\mathbf{q}) =
\delta_{\alpha\beta} - q_{\alpha}q_{\beta}/q^2$ (with $\alpha , \beta = x,y,z$), which is a projector on to
the plane perpendicular to the direction of $\mathbf{q}$,
and ${\boldsymbol \chi}(\mathbf{q},\mathbf{q};\omega)$ is the $3 \times 3$ spin susceptibility matrix. The poles of $S(\mathbf{q},\mathbf{\omega})$ occur at frequencies of magnons and Stoner excitations. This quantity is accessible from the linear-response theory, and in the following we will show how it can be computed in an efficient way using the Liouville-Lanczos approach to TDDFpT.

In the following Hartree atomic units will be used.

\subsection{Ground state}
\label{sec:GS}

In periodic solids, in the spin-polarized case the ground-state Kohn-Sham (KS) equations read~\cite{Note:notations_circ}:
\begin{equation}
\hat{H}^\circ \Psi_{n,\mathbf{k}}^\circ(\mathbf{r}) =
\varepsilon_{n,\mathbf{k}}^\circ \Psi_{n,\mathbf{k}}^\circ(\mathbf{r}) \,,
\end{equation}
where $n$ is the band index, $\mathbf{k}$ is the point in the first Brillouin zone (1BZ),
$\varepsilon_{n,\mathbf{k}}^\circ$ are the KS energies,
and $\Psi_{n,\mathbf{k}}^\circ(\mathbf{r})$ are the ground-state two-component KS spinor wave functions
\begin{align}
\Psi_{n,\mathbf{k}}^\circ(\mathbf{r})
=
\begin{pmatrix}
\psi^\circ_{n,\mathbf{k},1}(\mathbf{r}) \\
\psi^\circ_{n,\mathbf{k},2}(\mathbf{r})
\end{pmatrix}
\,,
\end{align}
where the subscripts ``1'' and ``2'' correspond to spin-up ($\uparrow$) and spin-down
($\downarrow$) components of the spinor, respectively. The ground-state $2 \times 2$
Hamiltonian $\hat{H}^\circ$ reads:
\begin{equation}
\hat{H}^\circ = -\frac{1}{2} \nabla^2 + \hat{V}^\circ_\mathrm{ext}
+ \hat{V}^\circ_\mathrm{H} + \hat{V}^\circ_\mathrm{XC} \,,
\label{eq:Hamiltonian_0}
\end{equation}
where the first term is the kinetic-energy operator, $\hat{V}^\circ_\mathrm{ext} = \hat{V}^\circ_\mathrm{loc} + \hat{V}^\circ_\mathrm{NL}$ is the external potential, defined as the sum of the local and non-local parts of the pseudopotential (PP), $\hat{V}^\circ_\mathrm{H}$ and $\hat{V}^\circ_\mathrm{XC}$ are the Hartree and exchange-correlation (XC) potentials, respectively. The last two operators in Eq.~\eqref{eq:Hamiltonian_0} depend on the $2 \times 2$ spin-charge density, which reads:
\begin{equation}
n^\circ_{\sigma\sigma'}(\mathbf{r}) = \sum_{n,\mathbf{k}} f_{n,\mathbf{k}} \,
\psi^{\circ *}_{n,\mathbf{k},\sigma}(\mathbf{r}) \psi^{\circ}_{n,\mathbf{k},\sigma'}(\mathbf{r}) \,,
\label{eq:spin-charge-density}
\end{equation}
where $\sigma$ and $\sigma'$ are the spin indices, $f_{n,\mathbf{k}}$ is the occupation factor which equals to 1 for occupied states and to 0 for empty states at zero temperature, hence $n$ runs over occupied states only, 
and $\mathbf{k}$ runs up to $N_\mathbf{k}$ points in 1BZ.
It is convenient to change variables and instead of working with $n^\circ_{\sigma\sigma'}(\mathbf{r})$ to use the charge density $n^\circ(\mathbf{r})$ and spin density (also called magnetization density) ${\boldsymbol m}^\circ(\mathbf{r})$, which are defined, respectively, as:
\begin{eqnarray}
n^\circ(\mathbf{r}) & = & \sum_\sigma n^\circ_{\sigma\sigma}(\mathbf{r}) \nonumber \\
& = & \sum_{n,\mathbf{k}} f_{n,\mathbf{k}} \,
\Psi_{n, \mathbf{k}}^{\circ \dagger}(\mathbf{r}) \, \Psi_{n, \mathbf{k}}^{\circ}(\mathbf{r}) \,,
\label{eq:charge_dens_0}
\end{eqnarray}
\begin{eqnarray}
{\boldsymbol m}^\circ(\mathbf{r}) & = & \mu_\mathrm{B} \sum_{\sigma \sigma'}
{\boldsymbol \sigma_{\sigma\sigma'}} \, n^\circ_{\sigma'\sigma}(\mathbf{r}) \nonumber \\
& = & \mu_\mathrm{B} \sum_{n,\mathbf{k}} f_{n,\mathbf{k}} \,
\Psi_{n, \mathbf{k}}^{\circ \dagger}(\mathbf{r}) \,
{\boldsymbol \sigma} \, \Psi_{n, \mathbf{k}}^{\circ}(\mathbf{r}) \,,
\label{eq:spin_dens_0}
\end{eqnarray}
where $\mu_\mathrm{B}$ is the Bohr magneton, and ${\boldsymbol \sigma} =
(\sigma_x, \sigma_y, \sigma_z)$ is the vector of Pauli matrices.
With these definitions, the Hartree potential $\hat{V}^\circ_\mathrm{H}$ in the coordinate
representation can be written as:
\begin{equation}
V^\circ_\mathrm{H}(\mathbf{r}) = \sigma^\circ \, v^\circ_\mathrm{H}(\mathbf{r}) \,,
\end{equation}
\begin{equation}
v^\circ_\mathrm{H}(\mathbf{r}) =
\int \frac{n^\circ(\mathbf{r}')}{|\mathbf{r} - \mathbf{r}'|} \,
d\mathbf{r}' \,,
\end{equation}
and the XC potential $\hat{V}^\circ_\mathrm{XC}$ in the coordinate representation is defined as:
\begin{equation}
V^\circ_\mathrm{XC}(\mathbf{r}) = \sigma^\circ v^\circ_\mathrm{XC}(\mathbf{r}) -
\mu_\mathrm{B} \, {\boldsymbol \sigma} \cdot {\boldsymbol b}^\circ_\mathrm{XC}(\mathbf{r}) \,,
\end{equation}
where $\sigma^\circ$ is the unit $2 \times 2$ matrix, and $v^\circ_\mathrm{XC}(\mathbf{r})$ and
${\boldsymbol b}^\circ_\mathrm{XC}(\mathbf{r})$ are the scalar and magnetic parts of the XC potential which are defined as:
\begin{align}
v^\circ_\mathrm{XC}(\mathbf{r}) & = \biggl. \frac{\delta E_\mathrm{XC}[n,{\boldsymbol m}]}{\delta n}
\biggr|_{\substack{\scriptscriptstyle n=n^\circ(\mathbf{r}) \\ \scriptscriptstyle {\boldsymbol m} = {\boldsymbol m}^\circ(\mathbf{r})}},
\\[4pt]
{\boldsymbol b}^\circ_\mathrm{XC}(\mathbf{r}) & =
-\biggl. \frac{\delta E_\mathrm{XC}[n,{\boldsymbol m}]}{\delta {\boldsymbol m}}
\biggr|_{\substack{\scriptscriptstyle n=n^\circ(\mathbf{r}) \\ \scriptscriptstyle {\boldsymbol m} = {\boldsymbol m}^\circ(\mathbf{r})}} \,,
\end{align}
where $E_\mathrm{XC}[n,{\boldsymbol m}]$ is the XC energy functional.

For the sake of convenience, let us rewrite Hamiltonian~\eqref{eq:Hamiltonian_0} as:
\begin{equation}
\hat{H}^{\circ} = \sigma^\circ \left[ -\frac{1}{2} \nabla^2 + \hat{v}^\circ_\mathrm{ext}
+ \hat{v}^\circ_\mathrm{H} + \hat{v}^\circ_\mathrm{XC} \right]
- \mu_\mathrm{B} \, {\boldsymbol \sigma} \cdot \hat{{\boldsymbol b}}^\circ_\mathrm{XC} \,,
\label{eq:Hamiltonian_0_minus}
\end{equation}
where we have used the notation $\hat{V}^\circ_\mathrm{ext} = \sigma^\circ \hat{v}^\circ_\mathrm{ext}$.

\subsection{Time-dependent density functional perturbation theory}

\subsubsection{General formulation}
\label{sec:general_formulation}

According to the Bloch theorem, the KS spinor wave functions can be written as:
\begin{equation}
\Psi^\circ_{n,\mathbf{k}}(\mathbf{r}) = \frac{1}{\sqrt{N_\mathbf{k}}} \,
e^{i\mathbf{k}\cdot\mathbf{r}} \, U^\circ_{n,\mathbf{k}}(\mathbf{r}) \,,
\label{eq:Psi_GS_Bloch}
\end{equation}
where $U^\circ_{n,\mathbf{k}}(\mathbf{r})$ are the lattice-periodic spinor functions, and the normalization factor $1/\sqrt{N_\mathbf{k}}$ is present because $U^\circ_{n,\mathbf{k}}(\mathbf{r})$ are taken to be orthonormalized in the primitive unit cell. We consider a system, initially in the ground state described by  Hamiltonian~\eqref{eq:Hamiltonian_0_minus}, perturbed by adiabatically switching on a time-dependent external potential. In the frequency domain, the perturbing potential can be decomposed into monochromatic components~\cite{Baroni:2001,Timrov:2013}: 
\begin{equation}
\tilde{V}^{\prime}_\mathrm{ext}(\mathbf{r}, \omega) =
\sum_\mathbf{q} e^{i\mathbf{q}\cdot\mathbf{r}} \,
\tilde{V}^{\prime}_{\mathrm{ext},\mathbf{q}}(\mathbf{r}, \omega) \,,
\label{eq:V_ext_q-decomposition}
\end{equation}
where $\tilde{V}^{\prime}_{\mathrm{ext},\mathbf{q}}(\mathbf{r}, \omega)$
is the lattice-periodic part. In the case of a magnetic perturbation it reads: 
\begin{equation}
\tilde{V}^{\prime}_{\mathrm{ext},\mathbf{q}}(\mathbf{r}, \omega) =
- \mu_\mathrm{B} \, {\boldsymbol \sigma} \cdot
\tilde{\boldsymbol b}'_{\mathrm{ext},\mathbf{q}}(\mathbf{r}, \omega) \,,
\label{eq:V_ext_q}
\end{equation}
where $\tilde{\boldsymbol b}'_{\mathrm{ext},\mathbf{q}}(\mathbf{r}, \omega)$ is the lattice-periodic part of the external magnetic field potential. A more generic form of the external perturbation is discussed in appendix~\ref{sec:general_pert}. The response of KS spinor wave functions can be correspondingly expressed as a linear combination of the responses to each monochromatic $\mathbf{q}$ component of the perturbing potential~\cite{Timrov:2013}:
\begin{equation}
\tilde{\Psi}'_{n,\mathbf{k}}(\mathbf{r},\omega) = \frac{1}{\sqrt{N_\mathbf{k}}} \,
\sum_\mathbf{q} e^{i(\mathbf{k+q})\cdot\mathbf{r}} \,
\tilde{U}'_{n,\mathbf{k+q}}(\mathbf{r},\omega) \,,
\label{eq:Phi_response_Bloch}
\end{equation}
where $\tilde{U}'_{n,\mathbf{k+q}}(\mathbf{r},\omega)$ are the lattice-periodic response spinor functions. Consequently, a similar decomposition can me made for the response charge and magnetization densities, and response Hartree-and-XC (HXC) potential. After such decompositions and by performing a linearization and Fourier transformation of the time-dependent KS equations, we can write the resonant and anti-resonant linear-response KS equations for individual monochromatic $\mathbf{q}$ components of the lattice-periodic quantities in the frequency domain as~\cite{Gorni:2016}:
\begin{widetext}
\begin{align}
\bigl( \hat{H}^{\circ}_\mathbf{k+q} - \varepsilon^\circ_{n,\mathbf{k}} - \omega \bigr) \,
\tilde{U}'_{n,\mathbf{k+q}}(\mathbf{r},\omega) + \, \hat{P}_{\mathbf{k+q}} \,
\hat{\tilde{V}}^{\prime}_{\mathrm{HXC},\mathbf{q}}(\omega) \,
U^\circ_{n,\mathbf{k}}(\mathbf{r})
& = - \hat{P}_{\mathbf{k+q}} \,
\hat{\tilde{V}}^{\prime}_{\mathrm{ext},\mathbf{q}}(\omega)
\, U^\circ_{n,\mathbf{k}}(\mathbf{r}) \,,
\label{eq:KS_eq_resonant_projected_q} 
\\
\bigl( \hat{H}^{\circ +}_\mathbf{k+q} - \varepsilon^\circ_{n,-\mathbf{k}} + \omega \bigr) \,
\hat{\mathrm{T}} \tilde{U}'_{n,-\mathbf{k}-\mathbf{q}}(\mathbf{r},-\omega) + \, \hat{P}^{+}_{\mathbf{k+q}} \,
\hat{\tilde{V}}^{\prime +}_{\mathrm{HXC},\mathbf{q}}(\omega) \,
\hat{\mathrm{T}} U^\circ_{n,-\mathbf{k}}(\mathbf{r})
&= - \hat{P}^{+}_{\mathbf{k+q}} \,
\hat{\tilde{V}}^{\prime +}_{\mathrm{ext},\mathbf{q}}(\omega) \,
\hat{\mathrm{T}} U^\circ_{n,-\mathbf{k}}(\mathbf{r}) \,.
\label{eq:KS_eq_antiresonant_projected_q}
\end{align}
Equation~\eqref{eq:KS_eq_antiresonant_projected_q} can be obtained from Eq.~\eqref{eq:KS_eq_resonant_projected_q} by changing the sign of $\omega$, $\mathbf{k}$, and $\mathbf{q}$, and by applying the time-reversal operator $\hat{\mathrm{T}} = i \sigma_y \hat{K}$, where $\hat{K}$ is the complex-conjugation operator. Here,
\begin{equation}
\hat{H}^\circ_\mathbf{k+q} =
\sigma^\circ \biggl[ -\frac{1}{2} \left[\nabla + i (\mathbf{k+q})\right]^2 +
\hat{v}^\circ_\mathrm{NL, \mathbf{k+q}} 
+ \, \hat{v}^\circ_\mathrm{loc} + \hat{v}^\circ_\mathrm{H} +
\hat{v}^\circ_\mathrm{XC} \biggr] -
\mu_\mathrm{B} \, {\boldsymbol \sigma} \cdot \hat{{\boldsymbol b}}^\circ_\mathrm{XC} \,,
\label{eq:H0_k_plus_q}
\end{equation}
\end{widetext}
whereas in Eq.~\eqref{eq:KS_eq_antiresonant_projected_q} we defined
$\hat{H}^{\circ +}_\mathbf{k+q} \equiv \hat{\rm T} \hat{H}^{\circ}_\mathbf{-k-q}  \hat{\rm T}^{-1}$, 
which can be shown to be equal to the operator in Eq.~\eqref{eq:H0_k_plus_q} with 
the opposite sign in the ground-state magnetic XC potential. We note that 
in Eq.~\eqref{eq:H0_k_plus_q} only the first two operators (kinetic term and non-local PP) depend on 
the current value of $\mathbf{k+q}$. 
In Eq.~\eqref{eq:KS_eq_resonant_projected_q}, $\hat{\tilde{V}}^{\prime}_{\mathrm{HXC},\mathbf{q}}(\omega)$ is the monochromatic $\mathbf{q}$ component of the response HXC potential, which reads:
\begin{equation}
\hat{\tilde{V}}^{\prime}_{\mathrm{HXC},\mathbf{q}}(\omega) = \sigma^\circ \, \hat{\tilde{v}}'_{\mathrm{H},\mathbf{q}}(\omega) + \sigma^\circ \, \hat{\tilde{v}}'_{\mathrm{XC},\mathbf{q}}(\omega) - \mu_\mathrm{B} \, {\boldsymbol \sigma} \cdot \hat{\tilde{\boldsymbol b}}'_{\mathrm{XC},\mathbf{q}}(\omega) \,,
\label{eq:V_HXC_resp_q}
\end{equation}
where
\begin{equation}
\tilde{v}'_{\mathrm{H},\mathbf{q}}(\mathbf{r},\omega) =
\int \frac{\tilde{n}'_\mathbf{q}(\mathbf{r}',\omega)}{|\mathbf{r}-\mathbf{r}'|} \,
e^{-i\mathbf{q}\cdot(\mathbf{r}-\mathbf{r}')} \, d\mathbf{r}' \,,
\label{eq:v_H_q}
\end{equation}
is the response Hartree potential in the coordinate representation, 
and $\hat{\tilde{v}}'_{\mathrm{XC},\mathbf{q}}(\omega)$ and 
$\hat{\tilde{\boldsymbol b}}'_{\mathrm{XC},\mathbf{q}}(\omega)$ are the response scalar and magnetic XC potentials, respectively, which in the coordinate representation within ALSDA read~\cite{Baroni:2001, Note:notation_Vxc}:
\begin{widetext}
  \begin{align}
    \tilde{v}'_{\mathrm{XC},\mathbf{q}}(\mathbf{r},\omega) & = \frac{\partial v_\mathrm{XC}}{\partial n} \biggr|_{n^\circ,{\boldsymbol m}^\circ} \tilde{n}'_\mathbf{q}(\mathbf{r},\omega) + \, \frac{\partial v_\mathrm{XC}}{\partial {\boldsymbol m}} \biggr|_{n^\circ,{\boldsymbol m}^\circ} \tilde{\boldsymbol m}'_\mathbf{q}(\mathbf{r},\omega) \,, \label{eq:v_XC_q} \\
    \tilde{\boldsymbol b}'_{\mathrm{XC},\mathbf{q}}(\mathbf{r},\omega) & = \frac{\partial {\boldsymbol b}_\mathrm{XC}}{\partial n} \biggr|_{n^\circ,{\boldsymbol m}^\circ} \tilde{n}'_\mathbf{q}(\mathbf{r},\omega) + \, \frac{\partial {\boldsymbol b}_\mathrm{XC}}{\partial {\boldsymbol m}} \biggr|_{n^\circ,{\boldsymbol m}^\circ} \tilde{\boldsymbol m}'_\mathbf{q}(\mathbf{r},\omega) \,. \label{eq:b_XC_q}
  \end{align}
From Eqs.~\eqref{eq:v_XC_q} and \eqref{eq:b_XC_q} we can see that there are mixed scalar and magnetic responses of $v_\mathrm{XC}$ and ${\boldsymbol b}_\mathrm{XC}$, which are coupled in a self-consistent way. As will be seen in the following, this allows us to compute the spin susceptibility directly by avoiding calculations of charge-charge responses and cross-terms spin-charge responses (see also appendix~\ref{sec:general_pert}). The response potentials in Eqs.~\eqref{eq:v_H_q} -- \eqref{eq:b_XC_q} are expressed in terms of the monochromatic $\mathbf{q}$ components of the response charge and magnetization densities, which read:
\begin{align}
    \tilde{n}'_\mathbf{q}(\mathbf{r},\omega) & =
\frac{1}{N_\mathbf{k}} \sum_{n, \mathbf{k}} \biggl[ f_{n,\mathbf{k}} \, U^{\circ \dagger}_{n,\mathbf{k}}(\mathbf{r}) \, \tilde{U}'_{n,\mathbf{k+q}}(\mathbf{r},\omega) +  f_{n,-\mathbf{k}} \left(\hat{\rm T}U^\circ_{n,-\mathbf{k}}(\mathbf{r}) \right)^{\dag} \hat{\mathrm{T}}\tilde{U}'_{n,-\mathbf{k}-\mathbf{q}}(\mathbf{r},-\omega) \biggr] \,,
\label{eq:charge_dens_resp_q} \\
    \tilde{\boldsymbol m}'_\mathbf{q}(\mathbf{r},\omega) & =
\frac{\mu_\mathrm{B}}{N_\mathbf{k}} \sum_{n, \mathbf{k}}
\biggl[ f_{n,\mathbf{k}} \, U^{\circ \dagger}_{n,\mathbf{k}}(\mathbf{r}) \, {\boldsymbol \sigma} \, \tilde{U}'_{n,\mathbf{k+q}}(\mathbf{r},\omega) -  f_{n,-\mathbf{k}} \left(\hat{\rm T}U^\circ_{n,-\mathbf{k}}(\mathbf{r}) \right)^{\dag} \, {\boldsymbol \sigma} \, \hat{\mathrm{T}}\tilde{U}'_{n,-\mathbf{k}-\mathbf{q}}(\mathbf{r},-\omega) \biggr] \,,
\label{eq:spin_dens_resp_q}
\end{align}
%
and satisfy the following relations~\cite{Note:density_property}:
$\tilde{n}^{\prime *}_{-\mathbf{q}}(\mathbf{r},-\omega) = \tilde{n}'_\mathbf{q}(\mathbf{r},\omega)$ and $\tilde{\boldsymbol m}^{\prime *}_{-\mathbf{q}}(\mathbf{r},-\omega) = \tilde{\boldsymbol m}'_\mathbf{q}(\mathbf{r},\omega)$. Using these properties it is easy to see that 
$ \hat{\tilde{V}}^{\prime +}_{\mathrm{HXC},\mathbf{q}}(\omega) \equiv
\hat{\rm T} \, \hat{\tilde{V}}^{\prime}_{\mathrm{HXC},-\mathbf{q}}(-\omega) \hat{\rm T}^{-1}$
is the operator of Eq.~\eqref{eq:V_HXC_resp_q} with the opposite sign in the response magnetic XC potential. The same applies for $\hat{\tilde{V}}^{\prime +}_{\mathrm{ext},\mathbf{q}}(\omega) 
\equiv \hat{\rm T} \, \hat{\tilde{V}}^{\prime}_{\mathrm{ext},-\mathbf{q}}(-\omega) \hat{\rm T}^{-1}$, which is the external perturbing potential~\eqref{eq:V_ext_q} with a reversed direction of the magnetic field.
Lastly, the operators $\hat{P}_{\mathbf{k+q}}$ and
$\hat{P}^{+}_{\mathbf{k+q}}$, appearing in Eqs.~\eqref{eq:KS_eq_resonant_projected_q} and \eqref{eq:KS_eq_antiresonant_projected_q}, respectively, are the projectors on to the empty-states manifold, and in the coordinate representation they read:
%
  \begin{align}
    P_{\mathbf{k+q}}(\mathbf{r},\mathbf{r}') & = \delta(\mathbf{r}-\mathbf{r}') -
\sum_{m} f_{m,\mathbf{k}+\mathbf{q}}\, U^\circ_{m,\mathbf{k+q}}(\mathbf{r}) \, U^{\circ \dagger}_{m,\mathbf{k+q}}(\mathbf{r}') \,, \label{eq:P_minus_q2} \\
    P^{+}_{\mathbf{k+q}}(\mathbf{r},\mathbf{r}') & =
\hat{\rm T} \, P_{-\mathbf{k}-\mathbf{q}}(\mathbf{r},\mathbf{r}')  \, \hat{\rm T}^{-1} \nonumber \\
    &  =  \delta(\mathbf{r}-\mathbf{r}') - \sum_{m} f_{m,-\mathbf{k}-\mathbf{q}}\left( \hat{\mathrm{T}} U^\circ_{m,-\mathbf{k}-\mathbf{q}}(\mathbf{r}) \right) \left( \hat{\mathrm{T}} U^{\circ}_{m,-\mathbf{k}-\mathbf{q}}(\mathbf{r}') \right)^\dagger \,. \label{eq:P_plus_q2}
  \end{align}
\end{widetext}
We stress that the projectors on to the empty-states manifold 
$\hat{P}_{\mathbf{k+q}}$ and $\hat{P}^{+}_{\mathbf{k+q}} $ are expressed in terms of the 
ground-state spinors $U^{\circ}_{m,\mathbf{k}+\mathbf{q}}$ and 
$U^{\circ}_{m,-\mathbf{k}-\mathbf{q}}$, respectively, which in turn refer to the 
occupied-states manifold, similarly to the static DFpT~\cite{Baroni:1987,Baroni:2001}. Therefore, no explicit reference to empty states is present in our formulation, \emph{i.e.} we avoid computationally expensive summations over empty states. 

In summary of this section, the linear-response problem is decoupled for individual monochromatic $\mathbf{q}$ components of the external magnetic perturbation, and is described by the resonant and anti-resonant linear-response KS equations~\eqref{eq:KS_eq_resonant_projected_q} and \eqref{eq:KS_eq_antiresonant_projected_q}, respectively. A generalization of the formalism to metals is shown in Appendix~\ref{sec:metals}.

\subsubsection{Quantum Liouville equation and spin susceptibility matrix}
\label{sec:quantum_Liouville_eq}

The resonant and anti-resonant linear-response KS equations~\eqref{eq:KS_eq_resonant_projected_q} and \eqref{eq:KS_eq_antiresonant_projected_q} can be equivalently expressed in terms of the quantum Liouville equation for the $2 \times 2$ response spin-charge density matrix operator $\hat{\tilde{\rho}}_{\mathbf q}'(\omega)$~\cite{Timrov:2013}:
\begin{equation}
( \omega - \hat{\mathcal{L}}_\mathbf{q} ) \cdot \hat{\tilde{\rho}}_{\mathbf q}'(\omega)
= [\hat{\tilde{V}}^{\prime}_\mathrm{ext,\mathbf{q}}(\omega), \hat{\rho}^\circ] \,,
\label{eq:Liouville_eq_1}
\end{equation}
where $\hat{\tilde{V}}^{\prime}_\mathrm{ext,\mathbf{q}}(\omega)$ is the external perturbing potential defined in Eq.~\eqref{eq:V_ext_q}, $\hat{\rho}^{\circ}$ is the unperturbed $2 \times 2$ spin-charge density matrix operator, and $\hat{\mathcal{L}}_{\mathbf q}$ is the Liouvillian superoperator, the action of which is defined as:
\begin{equation}
\hat{\mathcal{L}}_\mathbf{q} \cdot \hat{\tilde{\rho}}'_{\mathbf q}(\omega) \equiv
\left[\hat{H}^{\circ}, \hat{\tilde{\rho}}_{\mathbf q}'(\omega)\right] +
\left[\hat{\tilde{V}}^{\prime}_{\mathrm{HXC},\mathbf{q}}[\hat{\tilde{\rho}}_{\mathbf q}'(\omega)], \hat{\rho}^\circ\right] \,,
\label{eq:Liouvillian_def}
\end{equation}
where $\hat{\tilde{V}}^{\prime}_{\mathrm{HXC},\mathbf{q}}$ is the response HXC potential
[see Eq.~\eqref{eq:V_HXC_resp_q}].

The expectation value of the magnetization-density operator linearly induced by 
the external magnetic perturbing potential at a specific transferred momentum 
$\mathbf{q}$ and at a specific frequency $\omega$ can be defined as:
\begin{eqnarray}
\bigl\langle \hat{{\boldsymbol m}}_\mathbf{q}' \bigr\rangle_\omega & = &
\mathrm{Tr}[\hat{\boldsymbol m}_{\mathbf{q}}^\dagger \,
\hat{\rho}_{\mathbf{q}}'(\omega)] \nonumber \\
& = & \left( \hat{\boldsymbol m}_{\mathbf{q}}, (\omega - \hat{\mathcal{L}}_\mathbf{q})^{-1} \cdot
[\hat{\tilde{V}}^{\prime}_\mathrm{ext, \mathbf{q}}(\omega), \hat{\rho}^\circ] \right) \,,
\label{eq:m_expectation_value}
\end{eqnarray}
where with $(\cdot,\cdot)$ we indicate a scalar product in an operator space.
Using the following convention for the external perturbing potential~\cite{Note:ext_pot}
\begin{equation}
\hat{\tilde{V}}^{\prime}_\mathrm{ext,\mathbf{q}}(\omega) =
\hat{\boldsymbol m}_{\mathbf{q}} \cdot \hat{\tilde{\boldsymbol b}}'_{\mathrm{ext},\mathbf{q}}(\omega) \,,
\label{eq:V_to_m}
\end{equation}
we can rewrite the expectation value~\eqref{eq:m_expectation_value} as
\begin{equation}
\bigl\langle \hat{{\boldsymbol m}}_\mathbf{q}' \bigr\rangle_\omega =
 {\boldsymbol \chi}(\mathbf{q}, \mathbf{q}; \omega) \,
\hat{\tilde{\boldsymbol b}}'_{\mathrm{ext},\mathbf{q}}(\omega) \,,
\end{equation}
where ${\boldsymbol \chi}(\mathbf{q}, \mathbf{q}; \omega)$ is the $3 \times 3$ spin susceptibility matrix, which reads:
\begin{equation}
{\boldsymbol \chi}(\mathbf{q}, \mathbf{q}; \omega) =
\left( \hat{\boldsymbol m}_{\mathbf{q}}, (\omega - \hat{\mathcal{L}_\mathbf{q}})^{-1} \cdot
[\hat{\boldsymbol m}_{\mathbf{q}}, \hat{\rho}^\circ] \right) \,.
\label{eq:spin-density_susceptibility_2}
\end{equation}
The poles of this quantity mark the magnetic excitations of the system, and they allow to characterize the cross section of numerous magnetic spectroscopies, both bulk ones such as INSS [Eqs.~\eqref{eq:cross_section}--\eqref{eq:S_def}], or surface ones such as SPEELS~\cite{Gokhale:1992}.
It is worth noting that our formalism allows us to compute the whole $4 \times 4$ generalized susceptibility matrix which contains spin-spin [Eq.~\eqref{eq:spin-density_susceptibility_2}], charge-charge, spin-charge, and charge-spin couplings (see Appendix~\ref{sec:general_pert}).
Moreover, this is done in the general non-collinear framework, which is important in the presence of large spin-orbit coupling~\cite{Zakeri:2012} or in systems with complex non-collinear patterns in the ground state~\cite{Note:SOC}.

\subsection{Liouville-Lanczos approach}
\label{sec:LLapproach}

\subsubsection{Batch representation}
\label{sec:batch}

Equations~\eqref{eq:charge_dens_resp_q} and \eqref{eq:spin_dens_resp_q} show that
the response charge and magnetization densities are uniquely determined by the two sets of spinor wave functions $X_{\mathbf{q}} = \{ x_{n,\mathbf{k}+\mathbf{q}} \}$ and
$Y_{\mathbf{q}} = \{ y_{n,\mathbf{k}+\mathbf{q}} \}$,
which are called respectively upper and lower components of the {\it batch representation} (BR)
of the response spin-charge density matrix operator:
\begin{equation}
\hat{\tilde{\rho}}'_{\mathbf{q}} \xrightarrow{\mathrm{BR}}
\left(
\begin{array}{c}
X_{\mathbf{q}} \\ [5pt]
Y_{\mathbf{q}}
\end{array}
\right) =
\left(
\begin{array}{c}
\{ \tilde{U}'_{n,\mathbf{k}+\mathbf{q}}(\mathbf{r},\omega) \} \\ [5pt]
\{ \hat{\mathrm{T}}\tilde{U}'_{n,-\mathbf{k}-\mathbf{q}}(\mathbf{r},-\omega) \}
\end{array}
\right) \,.
\label{eq:batch_rho}
\end{equation}
This mapping can be formalized by defining BR of a generic operator $\hat{O}_\mathrm{\mathbf{q}}(\omega)$ as
\begin{align}
  \hat{O}_\mathrm{\mathbf{q}}(\omega)
  & \xrightarrow{\mathrm{BR}}\left(
\begin{array}{c}
O_{\mathbf{q}}^X \\[4pt]
O_{\mathbf{q}}^{Y}
\end{array}
\right) \nonumber \\
  & ~~~ = \quad \left(
  \begin{array}{c} \left\{ \hat{P}_{\mathbf{k}+\mathbf{q}} \,
  \hat{O}_\mathrm{\mathbf{q}}(\omega) \,
  U^\circ_{n,\mathbf{k}}(\mathbf{r}) \right\} \\ [6pt]
  \left\{ \hat{\mathrm{T}} \hat{P}_{-\mathbf{k}-\mathbf{q}} \,
  \hat{O}^{\dag}_\mathrm{\mathbf{q}}(\omega) \,
   U^\circ_{n,-\mathbf{k}}(\mathbf{r}) \right\}
\end{array}
\right) \,,
\label{eq:commutator_BR}
\end{align}
similarly to how it is done in Refs.~\cite{Rocca:2008, Malcioglu:2011}. 
Therefore, the commutator appearing on the right-hand side of 
Eq.~\eqref{eq:Liouville_eq_1} in BR will result in:
\begin{align}
  [\hat{\tilde{V}}^{\prime}_\mathrm{ext,\mathbf{q}}, \hat{\rho}^\circ]
  & \xrightarrow{\mathrm{BR}}\left(
\begin{array}{c}
V_{\mathbf{q}}^X \\[4pt]
V_{\mathbf{q}}^{Y}
\end{array}
\right) \nonumber \\
  & ~~~ = \quad \left(
  \begin{array}{c} \left\{ \hat{P}_{\mathbf{k}+\mathbf{q}} \,
  \hat{\tilde{V}}^{\prime}_\mathrm{ext,\mathbf{q}} \,
  U^\circ_{n,\mathbf{k}}(\mathbf{r}) \right\} \\ [6pt]
  \left\{ - \hat{P}^{+}_{\mathbf{k}+\mathbf{q}} \,
  \hat{\tilde{V}}^{\prime +}_\mathrm{ext,\mathbf{q}} \,
  \hat{\mathrm{T}} U^\circ_{n,-\mathbf{k}}(\mathbf{r}) \right\}
\end{array}
\right) \,.
\label{eq:commutator_BR}
\end{align}
Thus, the quantum Liouville equation~\eqref{eq:Liouville_eq_1}
[or equivalently Eqs.~\eqref{eq:KS_eq_resonant_projected_q} and
\eqref{eq:KS_eq_antiresonant_projected_q}] in BR takes the following form:
\begin{equation}
(\omega - \hat{\mathcal{L}_\mathbf{q}})
\left(
\begin{array}{c}
X_{\mathbf{q}} \\
Y_{\mathbf{q}}
\end{array}
\right) =
\left(
\begin{array}{c}
V_{\mathbf{q}}^X \\[4pt]
V_{\mathbf{q}}^{Y}
\end{array}
\right) \,,
\label{eq:Liouville_eq_BR}
\end{equation}
and the Liouvillian in BR reads:
\begin{equation}
\hat{\mathcal{L}}_\mathbf{q} \xrightarrow{\mathrm{BR}}
\left(
\begin{array}{cc}
  \mathcal{D}^{XX}_\mathbf{q} + \mathcal{K}^{XX}_\mathbf{q}  & \mathcal{K}^{XY}_\mathbf{q} \\ [6pt]
 -\mathcal{K}^{YX}_\mathbf{q}     &  -\mathcal{D}^{YY}_\mathbf{q} - \mathcal{K}^{YY}_\mathbf{q}
\end{array}
\right) \,,
\label{eq:Liouvillian_BR_q}
\end{equation}
where the actions of the superoperators, appearing in Eq.~\eqref{eq:Liouvillian_BR_q},
on the response batches are defined as:
\begin{widetext}
  \begin{align}
    \mathcal{D}^{XX}_\mathbf{q} X_\mathbf{q} & \equiv \left\{ (\hat{H}^{\circ}_{\mathbf{k+q}} - \varepsilon^\circ_{n,\mathbf{k}}) \, x_{n,\mathbf{k+q}} \right\} \,,
    \\
    \mathcal{D}^{YY}_\mathbf{q} Y_\mathbf{q} & \equiv \left\{ (\hat{H}^{\circ +}_{\mathbf{k+q}} - \varepsilon^\circ_{n,-\mathbf{k}}) \, y_{n,\mathbf{k+q}} \right\} \,,
    \\
    \mathcal{K}^{XX}_\mathbf{q} X_\mathbf{q} +\mathcal{K}^{XY}_\mathbf{q} Y_\mathbf{q} & \equiv \left\{ \hat{P}_{\mathbf{k+q}} \hat{\tilde{V}}^{\prime}_{\mathrm{HXC},\mathbf{q}}\bigl[\{x_{n,\mathbf{k+q}}\}, \{y_{n,\mathbf{k+q}}\}\bigr] \, U^\circ_{n,\mathbf{k}}(\mathbf{r}) \right\} \,,
    \\
    \mathcal{K}^{YX}_\mathbf{q} X_\mathbf{q} +\mathcal{K}^{YY}_\mathbf{q} Y_\mathbf{q} & \equiv \left\{ \hat{P}^{+}_{\mathbf{k+q}} \hat{\tilde{V}}^{\prime +}_{\mathrm{HXC},\mathbf{q}}\bigl[\{x_{n,\mathbf{k+q}}\}, \{y_{n,\mathbf{k+q}}\}\bigr] \, \hat{\mathrm{T}} U^\circ_{n,-\mathbf{k}}(\mathbf{r}) \right\} \,.
    \label{eq:liouvillian_blocks}
  \end{align}
Finally, from the expression for the expectation value of
$\hat{{\boldsymbol m}}_{\mathbf{q}}'$, Eq.~\eqref{eq:m_expectation_value},
we  can see that we need to represent $\hat{\boldsymbol m}_{\mathbf{q}}$ in BR.
Formally we can write:
\begin{equation}
\hat{\boldsymbol m}_\mathbf{q} \xrightarrow{\rm BR}
\left(
\begin{array}{c}
{\boldsymbol m}_{\mathbf{q}}^{X} \\ [5pt]
{\boldsymbol m}_{\mathbf{q}}^{Y}
\end{array}
\right) =
\left(
\begin{array}{c}
\{ \hat{P}_{\mathbf{k}+\mathbf{q}} \, \hat{\boldsymbol m}_\mathbf{q} U^\circ_{n,\mathbf{k}}(\mathbf{r}) \} \\ [5pt]
\{ \hat{P}^{+}_{\mathbf{k}+\mathbf{q}} \, \hat{\boldsymbol m}_\mathbf{q} \hat{\mathrm{T}} U^\circ_{n,-\mathbf{k}}(\mathbf{r}) \}
\end{array}
\right) \,.
\label{eq:m_BR}
\end{equation}
Therefore, using Eq.~\eqref{eq:V_to_m} in Eq.~\eqref{eq:commutator_BR}, and using 
Eqs.~\eqref{eq:Liouvillian_BR_q} -- \eqref{eq:m_BR}, we can write the spin
susceptibility matrix~\eqref{eq:spin-density_susceptibility_2} in BR as:
\begin{equation}
{\boldsymbol \chi}(\mathbf{q}, \mathbf{q}; \omega) \xrightarrow{\mathrm{BR}}
\left( ({\boldsymbol m}_{\mathbf{q}}^{X}, {\boldsymbol m}_{\mathbf{q}}^{Y})^\top, (\omega -
\hat{\mathcal{L}}_\mathbf{q})^{-1} \cdot
({\boldsymbol m}_{\mathbf{q}}^{X}, - {\boldsymbol m}_{\mathbf{q}}^{Y})^\top \right) \,,
\label{eq:spin-density_susceptibility_3}
\end{equation}
\end{widetext}
which can be efficiently computed using iterative algorithms, as explained in the next section.

It is worth noting that due to the lack of time-reversal symmetry,
it is not useful to make a rotation of the batches as it was done for other
spectroscopies~\cite{Rocca:2008, Timrov:2013}.

\subsubsection{Lanczos algorithm}
\label{sec:Lanczos_algorithm}

The spin susceptibility matrix in the batch representation~\eqref{eq:spin-density_susceptibility_3}
is well suited to be computed using iterative algorithms, 
which allows us to avoid computationally
expensive inversion of the Liouvillian. Popular methods are
the Lanczos recursive biorthogonalization algorithm and the Davidson diagonalization
algorithm~\cite{Ge:2014}. Here we will use the non-Hermitian Lanczos recursive
biorthogonalization algorithm, the details of which
can be found in Refs.~\cite{Saad:2003, Rocca:2008,Baroni:2012}.

Let us consider a generic function $g(\omega)$ which is defined as the off-diagonal element
of the resolvent of the operator $\hat{L}$ as:
\begin{equation}
  g(\omega) = \left( u, (\omega - \hat{L})^{-1} v \right) ,
  \label{eq:off_diag_matrix_element_general}
\end{equation}
where $\hat{L}$ is the $P\times P$ non-Hermitian matrix, and $u$ and $v$ are
generic $P$-dimensional arrays. To this end we define two sets of
\emph{Lanczos vectors}, $\{v_i\}$ and $\{u_i\}$, through the recursive relations:
\begin{align}
  \beta_{i+1} \, v_{i+1} & = \hat{L} \, v_i - \alpha_i \, v_i - \gamma_i \, v_{i-1} ,
  \label{eq:Lanczos_chain_1} \\
  \gamma_{i+1} \, u_{i+1} & = \hat{L}^\dagger \, u_i - \alpha_i \, u_i - \beta_i \, u_{i-1} \,,
  \label{eq:Lanczos_chain_2}
\end{align}
where we define $u_0=v_0=0$, $u_1=v_1=v$, and
$\alpha_i$, $\beta_i$, and $\gamma_i$ are the \emph{Lanczos coefficients}. 
The Lanczos vectors satisfy the biorthogonality condition $(u_i,v_j)=\delta_{ij}$.
The set of vectors and coefficients generated through the recursion relations
\eqref{eq:Lanczos_chain_1} -- \eqref{eq:Lanczos_chain_2} is called
the \emph{Lanczos chain}. The Lanczos coefficients are computed at every step
of the Lanczos recursion. 
Equations~\eqref{eq:Lanczos_chain_1} and \eqref{eq:Lanczos_chain_2} show that $\hat{L}$ and $\hat{L}^\dagger$ must be applied at each Lanczos iteration, which thus require four times as many Hamiltonian builds as in a ground-state calculation; this factor can be brought down to just two by exploiting the pseudo-Hermiticity of the Liouvillian~\cite{Gruning:2011, Ge:2014}.
If we call $\bar{v}$ and $\bar{u}$ the vectors on the 
right-hand side of Eqs.~\eqref{eq:Lanczos_chain_1} and \eqref{eq:Lanczos_chain_2}, 
respectively, the Lanczos coefficients are defined as~\cite{Rocca:2008}:
\begin{align}
\alpha_i     & = ( u_i, \hat{L} \, v_i ) \,,
\label{eq:alpha}
\\[6pt]
\beta_{i+1}  & = \sqrt{ | \left( \bar{v}, \bar{u} \right) | } \,,
\label{eq:beta}
\\[6pt]
\gamma & = \mathrm{sign}\big[ \left( \bar{v}, \bar{u} \right) \big] \, \beta_{i+1}
\, .
\label{eq:gamma}
\end{align}
The Lanczos vectors thus generated have the property that they provide a tridiagonal
representation of $\hat{L}$. More specifically, if we define
the $P\times M$ matrices $^{\scriptscriptstyle M\!}U =
\{ u_1, u_2, \ldots, u_{\scriptscriptstyle M} \}$ and
$^{\scriptscriptstyle M\!}V =
\{ v_1, v_2, \ldots, v_{\scriptscriptstyle M} \}$, where $M$ is the number of
Lanczos iterations, one has:
\begin{equation}
\left(^{\scriptscriptstyle M\!}U\right)^\dagger \hat{L} \,\, ^{\scriptscriptstyle M\!}V = \,^{\scriptscriptstyle M\!}T ,
\label{eq:L_to_T_tridiag}
\end{equation}
where $^{\scriptscriptstyle M\!}T$ is the tridiagonal matrix
\begin{equation}
^{M\!}T = \left(\begin{array}{ccccc}
\alpha_1 & \gamma_2  &    0     & \ldots   &  0       \\
\beta_2  & \alpha_2  & \gamma_3 &   0      &  \vdots  \\
0        & \beta_3   & \alpha_3 & \ddots   &  0       \\
\vdots   &    0      & \ddots   & \ddots   & \gamma_{\scriptscriptstyle M} \\
0        & \ldots    &    0     & \beta_{\scriptscriptstyle M}  & \alpha_{\scriptscriptstyle M}
\end{array}\right) .
\label{eq:tridiagonal_matrix}
\end{equation}
In this representation, the matrix element of
Eq.~\eqref{eq:off_diag_matrix_element_general} can be expressed as~\cite{Rocca:2008}:
\begin{equation}
  g(\omega) \simeq \left ( ^{\scriptscriptstyle M\!}z , \left ( \omega \, ^{\scriptscriptstyle M\!}I  - \,
  ^{\scriptscriptstyle M\!}T \right )^{-1} \cdot \, ^{\scriptscriptstyle M\!}e_1 \right ) ,
\label{eq:resolvent_g}
\end{equation}
where $^{\scriptscriptstyle M\!}e_1 = \{1,0,\ldots,0\}$, and $^{\scriptscriptstyle M\!} z$ is the $M$-dimensional
vector defined as:
\begin{equation}
^{\scriptscriptstyle M\!} z =  \, \left( ^{\scriptscriptstyle M\!}V \right )^\dagger u \,.
\label{eq:zeta_coef_for_g}
\end{equation}
The right-hand side of Eq.~(\ref{eq:resolvent_g}) can be conveniently
computed by solving, for any given value of $\omega$, the equation:
\begin{equation}
\left( \omega \, ^{\scriptscriptstyle M\!}I - \, ^{\scriptscriptstyle M\!}T \right) {^{\scriptscriptstyle M\!}x} = \, ^{\scriptscriptstyle M\!}e_1 ,
\label{eq:eta_post_processing_for_g}
\end{equation}
and calculating the scalar product:
\begin{equation}
g(\omega) = \left ( {^{\scriptscriptstyle M\!} z} , {^{\scriptscriptstyle M\!} x} \right ) .
\label{eq:resolvent_Liouvillian_Lanczos_2_for_g}
\end{equation}
The vector $^{\scriptscriptstyle M\!} z$, Eq.~(\ref{eq:zeta_coef_for_g}), can be computed on
the fly during the Lanczos recursion, through the relation $z_i=\left(v_i,u\right )$.

In practice, the procedure outlined above is
performed in two steps. In the first step, which is by far the most
time consuming, one generates the tridiagonal matrix $^{\scriptscriptstyle M\!}T$,
Eq.~(\ref{eq:tridiagonal_matrix}), and the vector $^{\scriptscriptstyle M\!}z$,
Eq.~(\ref{eq:zeta_coef_for_g}). The strength of the Lanczos algorithm for
frequency-independent XC kernels is precisely due to the fact that
the tridiagonalization is done independently of the frequency.
In the second step (post-processing), $g(\omega)$ is
calculated using Eq.~(\ref{eq:resolvent_Liouvillian_Lanczos_2_for_g})
upon the solution of Eq.~(\ref{eq:eta_post_processing_for_g}), for
different frequencies $\omega$. In practice, a small imaginary part
$\epsilon$ is added to the frequency argument, $\omega \rightarrow
\omega + i \, \epsilon$, so as to regularize the function
$g(\omega)$. Setting $\epsilon$ to a non-zero value
amounts to broadening each individual spectral line or,
alternatively, to convoluting the function $g(\omega)$ with a
Lorentzian. Because of the tridiagonal form of the matrix $^{\scriptscriptstyle M\!}T$
the second step is computationally inexpensive.
The Lorentzian broadening $\epsilon$ of the magnetic spectrum can be easily changed 
(this might be useful \emph{e.g.} when comparing the theoretical spectrum with the experimental one
and adjusting the broadening of the former to better fit the latter)
by simply re-doing the post-processing calculation at a negligible cost. 
This is an important advantage with respect to other methods that require 
to fix the broadening
at the very beginning of the calculation with no possibility to change it at the end.
Lastly, different responses to a same
perturbation can be computed simultaneously from a same Lanczos
recursion, by computing different $z$ vectors on the fly.

The convergence of the computed magnetic excitation spectrum with respect to the
length of the Lanczos chains depends on the spectral range: the lower the frequency is,
the faster the convergence is. Therefore, magnon peaks in the spectrum
converge faster than the the Stoner continuum.

\section{Validation}
\label{sec:validation}

The Liouville-Lanczos approach within TDDFpT to model magnetic excitations has been implemented in the \QE\, package~\cite{Giannozzi:2009, Giannozzi:2017} and is scheduled to be distributed in one of its future releases. We now proceed to validate it by calculating the spin susceptibility~\eqref{eq:spin-density_susceptibility_2} for bulk
ferromagnetic bcc iron and fcc nickel, for which several TDDFT and MBPT studies exist together
with the experimental data.

\subsection{Technical details}
 
All the calculations for bulk Fe and bulk Ni have been performed using ALSDA to the XC functional. We have used norm-conserving pseudopotentials from the
PseudoDojo library~\cite{Setten:2018, DojoPP:2018} and the experimental lattice parameters
$a=2.86$~\AA \, for Fe~\cite{Basinski:1955} and $a=3.52$~\AA \, for Ni~\cite{Bandyopadhyaya:1977}. Kohn-Sham spinor wave functions were expanded in plane waves (PW) up to a kinetic energy cutoff of 70~Ry for Fe and 60~Ry for Ni, while the charge and magnetization densities and potentials were expanded in PWs with the cutoff 4 times larger than that for wave functions. For both Fe and Ni the first Brillouin zone has been sampled with a
uniform $\mathbf{k}$ point mesh centered at the $\Gamma$ point of size $36\times 36\times 36$,
and we have used a Gaussian smearing technique with a broadening parameter of 5~mRy.
With these parameters, the ground state magnetization (aligned along the $z$ axis) results in
$2.17$\,$\mu_{\rm B}$ and $0.62$\,$\mu_{\rm B}$ per atom for Fe and Ni, respectively,
which is in good agreement with the experimental values of $2.22$\,$\mu_{\rm B}$~\cite{Crangle:1971} and $0.60$\,$\mu_{\rm B}$~\cite{Iota:2007}, respectively.
The magnon spectra have been convoluted with a Lorentzian function with a broadening parameter
of 0.5~mRy.

\subsection{Lanczos coefficients and convergence of magnon spectra}

In this section we analyze the behaviour of the Lanczos and $z$ coefficients and the convergence of a magnon spectrum with respect to the number of Lanczos iterations (see Sec.~\ref{sec:Lanczos_algorithm}) in the case of bulk Fe for the finite transferred momentum $\mathbf{q} = \frac{2\pi}{a} \, (0.2, 0.2, 0.0)$. The convergence studies for bulk Ni are qualitatively similar, hence we will not discuss them here.

\begin{figure}[t]
\centering
\includegraphics[width=0.97\columnwidth]{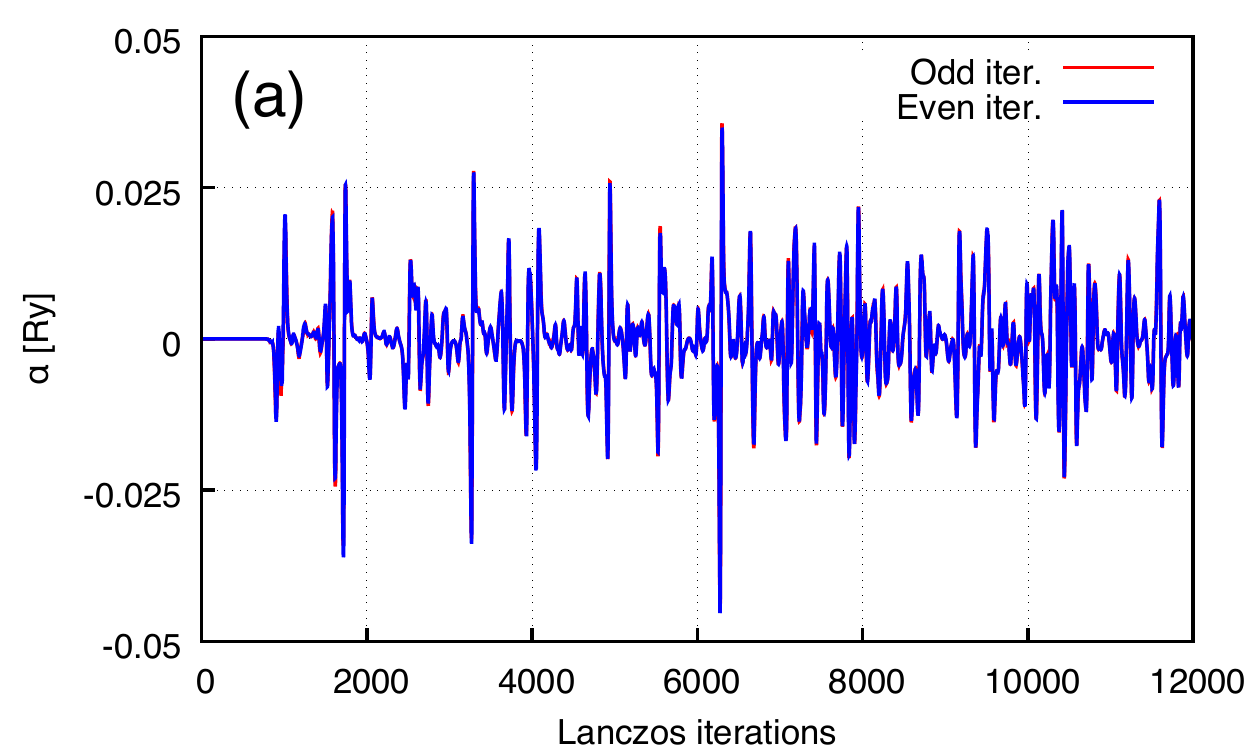}
\includegraphics[width=0.95\columnwidth]{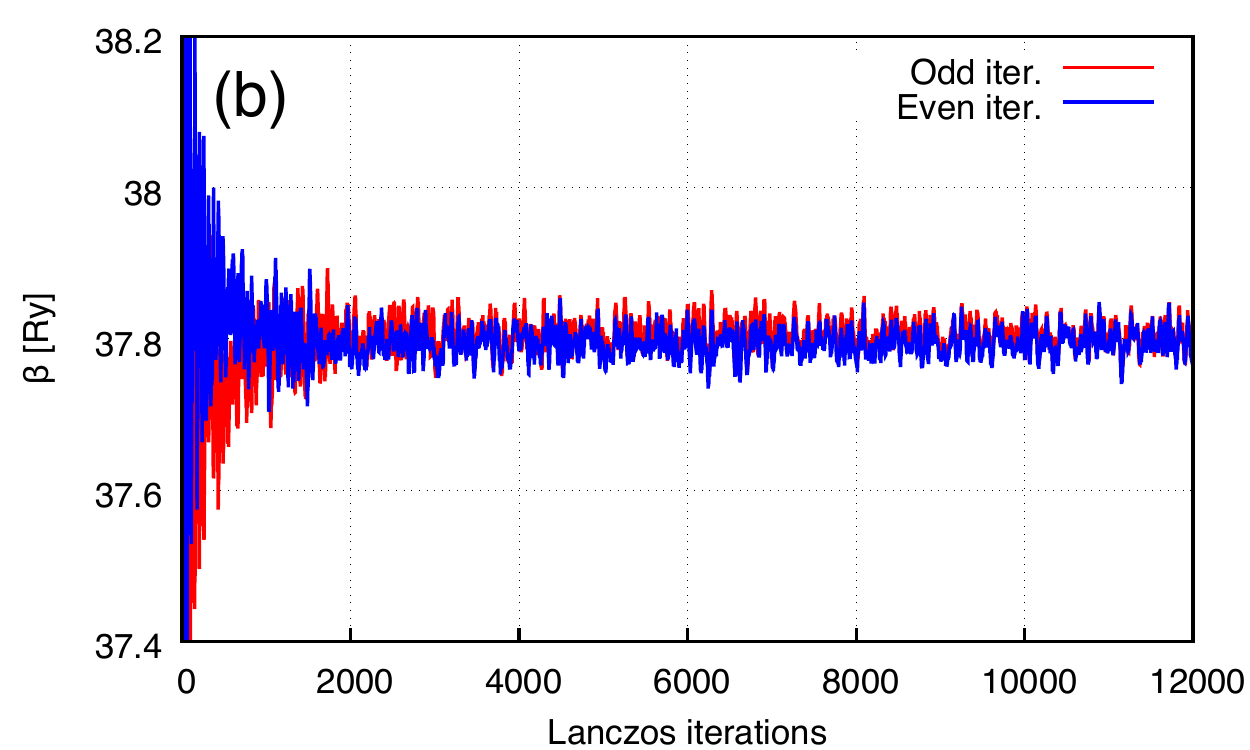}
\includegraphics[width=0.95\columnwidth]{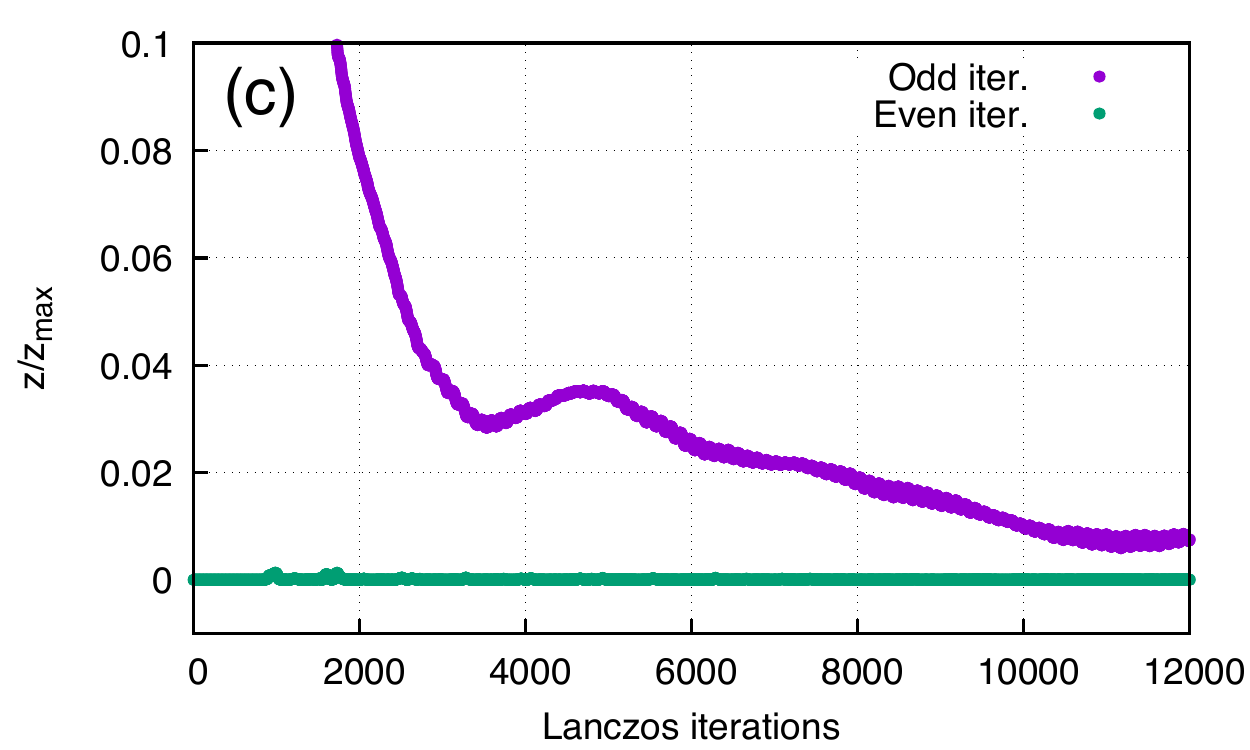}
\caption{
Behaviour of coefficients of the Lanczos algorithm as a function of the number of odd and even iterations for Fe. 
(a)~The $\alpha$ coefficients, defined in Eq.~\eqref{eq:alpha}, 
(b)~the $\beta$ coefficients, defined in Eq.~\eqref{eq:beta}, (c) $z/z_\mathrm{max}$, where $z$ coefficients are defined in Eq.~\eqref{eq:zeta_coef_for_g} and $z_\mathrm{max}$ is the maximum value of $z$. 
} 
\label{fig:lancz_coeff}
\end{figure}

We find that the $\alpha$ coefficients oscillate around zero; these oscillations are about three orders of magnitude smaller than the average value of $\beta$ (and $\gamma$), 
as it can be seen in Fig.~\ref{fig:lancz_coeff}. 
As was pointed out in other works which use the Lanczos algorithm~\cite{Rocca:2008, Malcioglu:2011}, $\beta$ coefficients oscillate around the energy equal approximatively to the half of the kinetic energy cutoff, whereas the difference of $\beta$'s at even and odd iterations corresponds roughly to twice the lowest excitation energy.
Indeed, here we find a similar trend, namely the $\beta$ coefficients oscillate around 37.8~Ry [see Fig.~\ref{fig:lancz_coeff}~(b)] which is roughly $70/2 = 35$~Ry, and the difference between the averages of $\beta$'s at even and odd iterations gives 220~meV which equals to twice the lowest excitation energy, which is in this case is the magnon energy of $\approx 120$~meV (see Sec.~\ref{sec:magnon_dispersion}). It is worth noting that in general there may be instabilities in the behaviour of $\beta$ coefficients~\cite{Rocca:2008, Malcioglu:2011} (which though do not influence the final spectra), however we did not observe any instabilities in the case presented here. Figure~\ref{fig:lancz_coeff}~(c) shows the evolution of $z$ coefficients defined in Eq.~\eqref{eq:zeta_coef_for_g}. It can be seen that $z$'s at even iterations are essentially zero 
, while $z$'s at odd iterations are non-zero and they decrease non-monotonically with the number of Lanczos iterations. 

\begin{figure}[t]
\centering
\includegraphics[width=0.9\columnwidth]{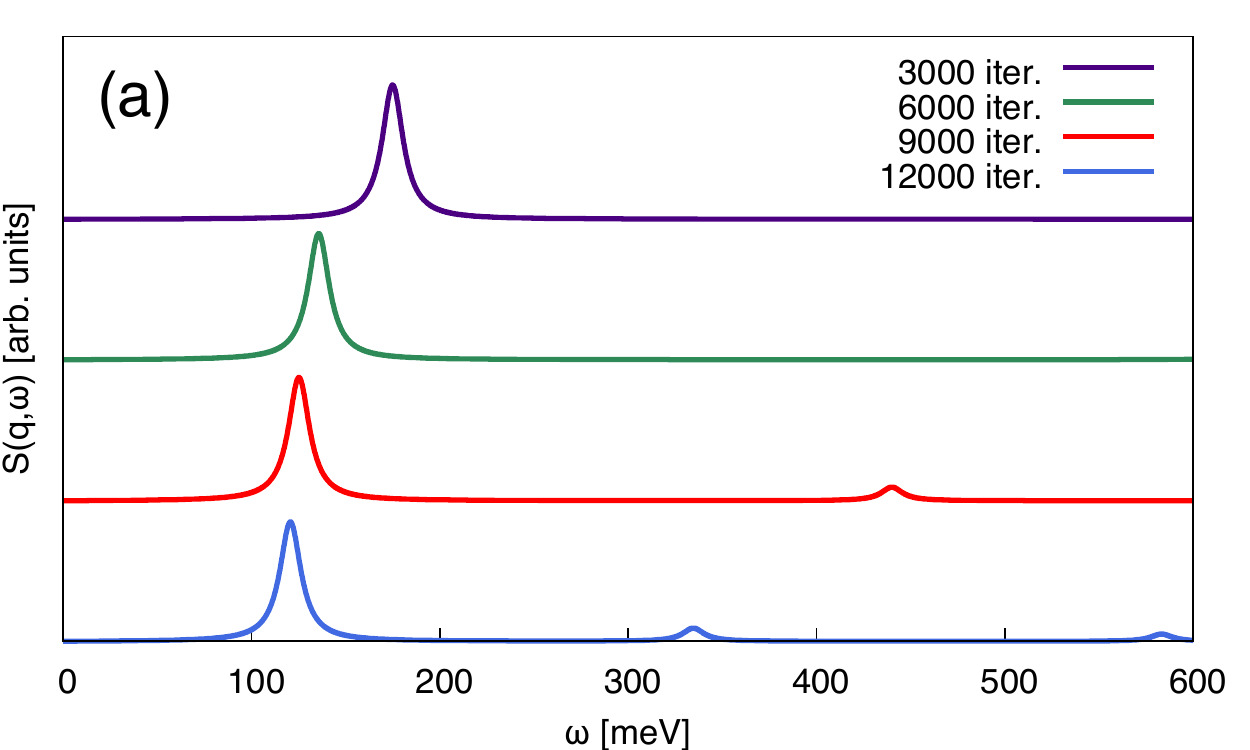}
\includegraphics[width=0.9\columnwidth]{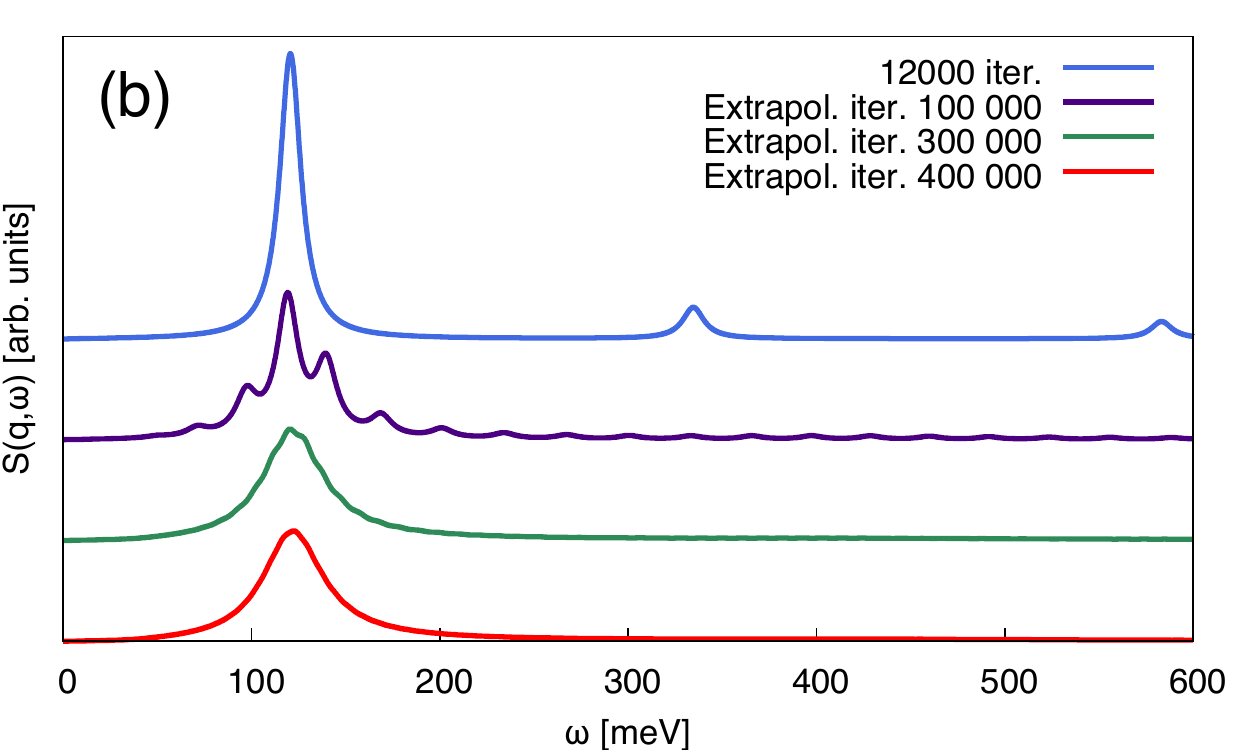}
\caption{
Convergence of the magnon peak in Fe for $\mathbf{q} = \frac{2\pi}{a} (0.2,0.2,0.0)$ which is a pole of the $S(\mathbf{q},\omega)$ function. (a)~No extrapolation is used. (b) The extrapolation technique is used except for the topmost spectrum (blue) which has been obtained with 12000 iterations without the extrapolation. In (a) and (b) the spectra have been shifted vertically for clarity.}
\label{fig:conv_no_extrap}
\end{figure}

In the following we discuss the convergence of the magnon spectra with respect to the number of Lanczos iterations. Magnons and Stoner excitations are the poles of
$S(\mathbf{q},\omega)$ [see Eq.~\eqref{eq:S_def}], which is directly related to the spin susceptibility matrix defined in Eq.~\eqref{eq:spin-density_susceptibility_2}. In Fig.~\ref{fig:conv_no_extrap}~(a) we show $S(\mathbf{q},\omega)$ for Fe, which was evaluated after performing a Lanczos calculation and determining Lanczos and $z$ coefficients up to a certain number of iterations. 
It can be seen that the magnon peak appears around 175~meV when computed after 3000 Lanczos iterations, and it shifts to smaller energies by further increasing the number of iterations. In addition, one can see smaller intensity peaks, \emph{e.g.} around 445~meV when computed after 9000 iterations, which shift largely during the Lanczos recursion. These peaks are due to Stoner excitations, and their position is very sensitive to the number of Lanczos iterations, indicating their slow convergence. Indeed, as was mentioned in Sec.~\ref{sec:Lanczos_algorithm}, the higher-energy portions of the spectra (i.e. Stoner excitations) tend to converge more slowly than the lower-energy ones (i.e. magnons) with respect to the number of Lanczos iterations. 
This problem can be overcome using the extrapolation technique for the Lanczos and $z$ coefficients~\cite{Rocca:2008}, which is a computationally inexpensive operation (negligible with respect to the cost of the Lanczos recursion calculation).
The main observation is that $z$ coefficients decrease with the Lanczos iterations and at some point they become very small [see Fig.~\ref{fig:lancz_coeff}~(c)]. 
Therefore, after a certain number of iterations $M_0$ -- after which $z$'s can be considered to be vanishing -- the spectrum is completely determined by $\alpha$, $\beta$
and $\gamma$ coefficients. These coefficients, in turn, can be extrapolated after $M_0$, because they do not show large variations but instead they oscillate around a certain number, as was explained above. 
Thus, setting the Lanczos coefficients to their respective averages for $M>M_0$ is an approximation which allows us to speed up considerably the convergence of the magnon spectrum, without the loss of accuracy. 
It is worth noting that the choice of $M_0$ is rather arbitrary, and in practice one has to perform convergence tests (which are computationally very cheap) with respect to this parameter.
In the case considered here, $z$ coefficients can be
considered to be equal to zero after $M_0 = 12000$ Lanczos iterations, where $z$ starts oscillating slightly below $0.01 \, z_{\rm max}$, as can be seen in Fig.~\ref{fig:lancz_coeff}~(c).
The magnon spectra computed using the extrapolation technique starting from 12000 Lanczos iterations are shown in Fig.~\ref{fig:conv_no_extrap}~(b). 
In order to obtain a converged spectrum it is necessary to extrapolate the Lanczos coefficients up to a few hundred thousands iterations, a value that would be unfeasible to reach explicitly via the Lanczos recursion without extrapolation, due to the large computational cost and the loss of stability of the algorithm due to increase of the numerical noise. 
The wiggles appearing \emph{e.g.} after the extrapolation up to 100000 do not have any physical meaning: they are inherent to the current approach -- at convergence no wiggles must be present (for example, after the extrapolation up to 400000 the spectrum is smooth with no wiggles). Moreover, in order to check the convergence of the magnon peak, we extrapolated the spectra also after $M_0 = 14000$ and $M_0 = 16000$ Lanczos iterations, and we did not observe any substantial changes. Finally, we note that the extrapolation of Lanczos coefficients does not alter the position of the magnon peak -- because it is already converged after 12000 Lanczos iterations without the extrapolation -- but it damps the magnon peak (decrease its intensity and increase the width) by bringing to convergence the Stoner continuum. 

All in all, the rather large number of Lanczos iterations necessary to reach convergence is known to be related to the condition number of the Liouvillian, \emph{i.e.} to the ratio between its maximum and minimum absolute eigenvalues. The minimum eigenvalue is the minimum excitation energy, while in a PW representation, the maximum eigenvalue is of the order of kinetic-energy cutoff. For magnetic excitations the condition number may be particularly large because magnetic excitations are in the meV range, whereas first-row transition metals,  usually responsible for magnetism, typically require rather large cutoffs.

In the case of charge excitations (plasmons), the Liouville-Lanczos approach has proved to be more convenient than the Sternheimer one, when the latter is used to compute the spectrum for more than 1--2 dozen frequencies~\cite{Motornyi:2018}. For magnetic excitations, the comparison may not be as favorable, due to a larger condition number of the Liouvillian in this case, as mentioned above. This condition number can likely be considerably reduced using a number of techniques (which we do not discuss in this work) thus reducing the number of Lanczos iterations.

\subsection{Discussion}
\label{sec:magnon_dispersion}

\begin{figure}[t!]
\centering
\hspace{4mm}
\includegraphics[width=0.9\columnwidth]{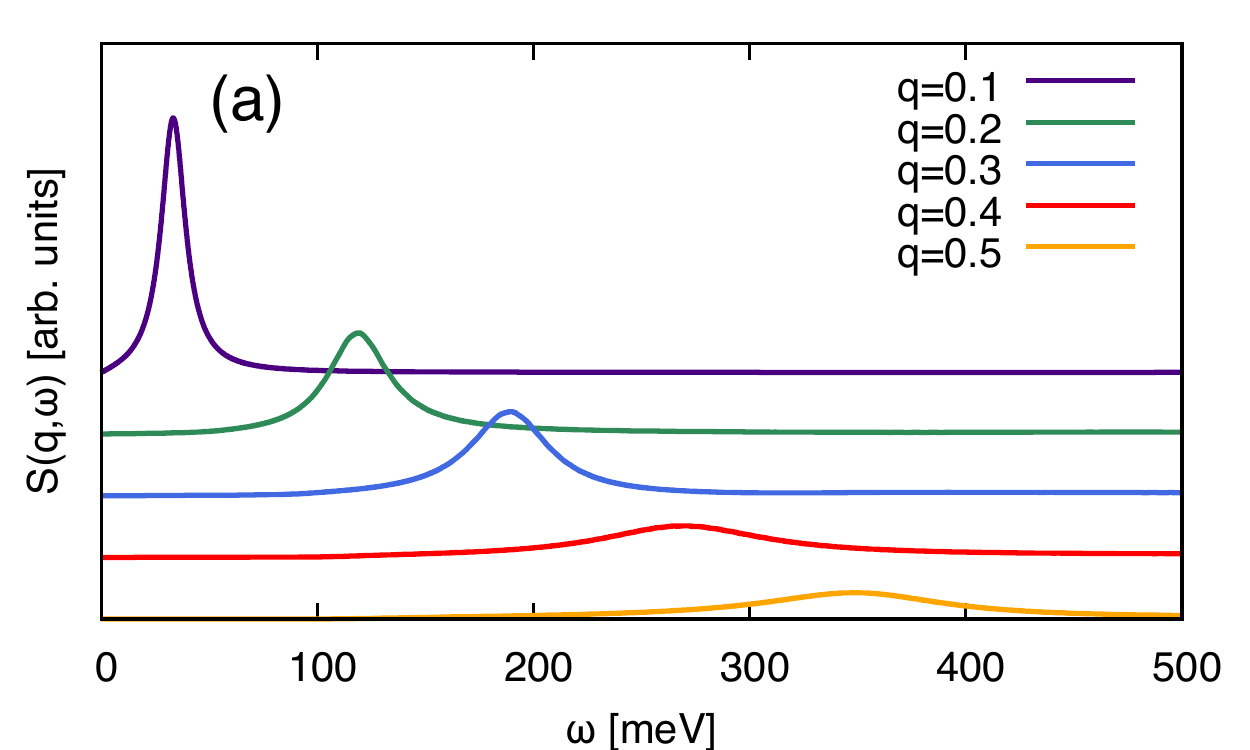}
\includegraphics[width=0.98\columnwidth]{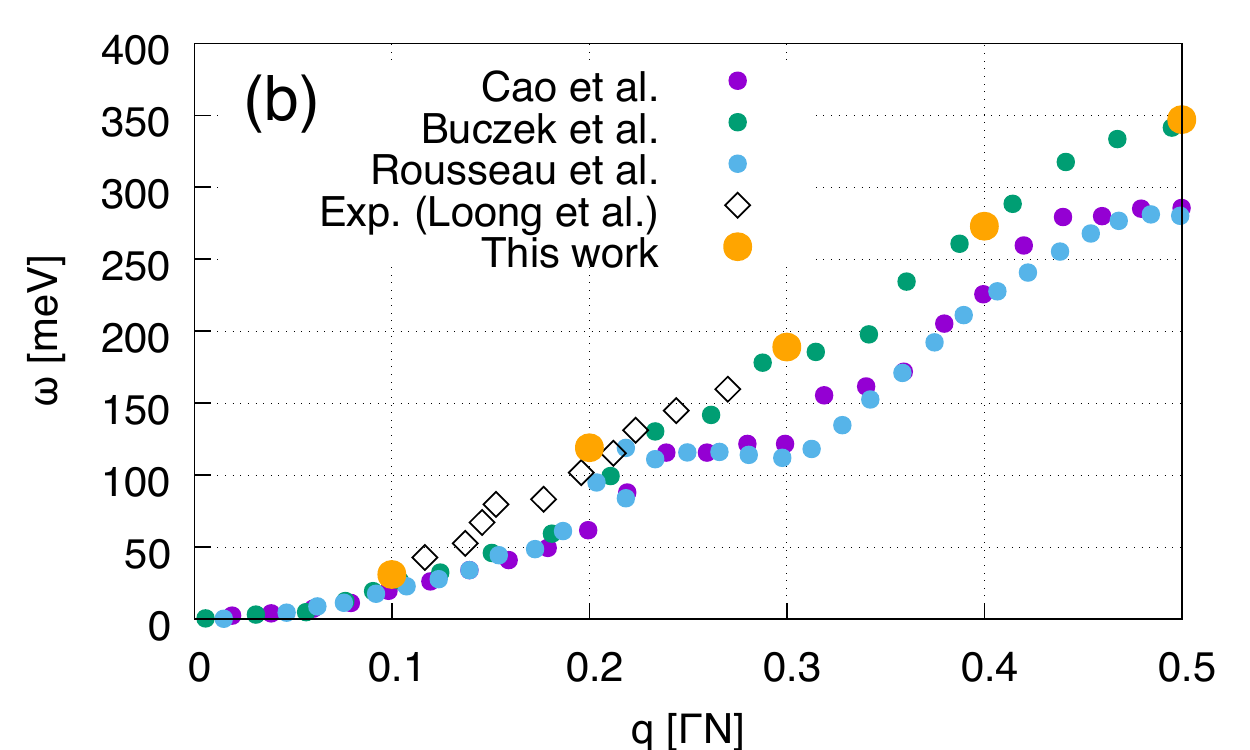}
\caption{
Magnon spectra and dispersion in Fe along the $\Gamma$-N direction in the Brillouin zone. (a)~Excitation spectrum $S(\mathbf{q},\omega)$ at several values of the transferred momentum $\mathbf{q} = \frac{2\pi}{a} (q,q,0)$. (b)~Comparison of the magnon dispersion as obtained in this work, in other theoretical works (Cao et al.~\cite{Cao:2018}, Buczek et al.~\cite{Buczek:2011b}, Rousseau et al.~\cite{Rousseau:2012}), and as measured in the INSS experiment at 10~K (Loong et al.~\cite{Loong:1984}). Each point in (b) represents the position of the maximum of $S(\mathbf{q},\omega)$ in (a).}
\label{fig:Fe_magnon}
\end{figure}

\begin{figure}[t!]
\centering
\hspace{4mm}
\includegraphics[width=0.9\columnwidth]{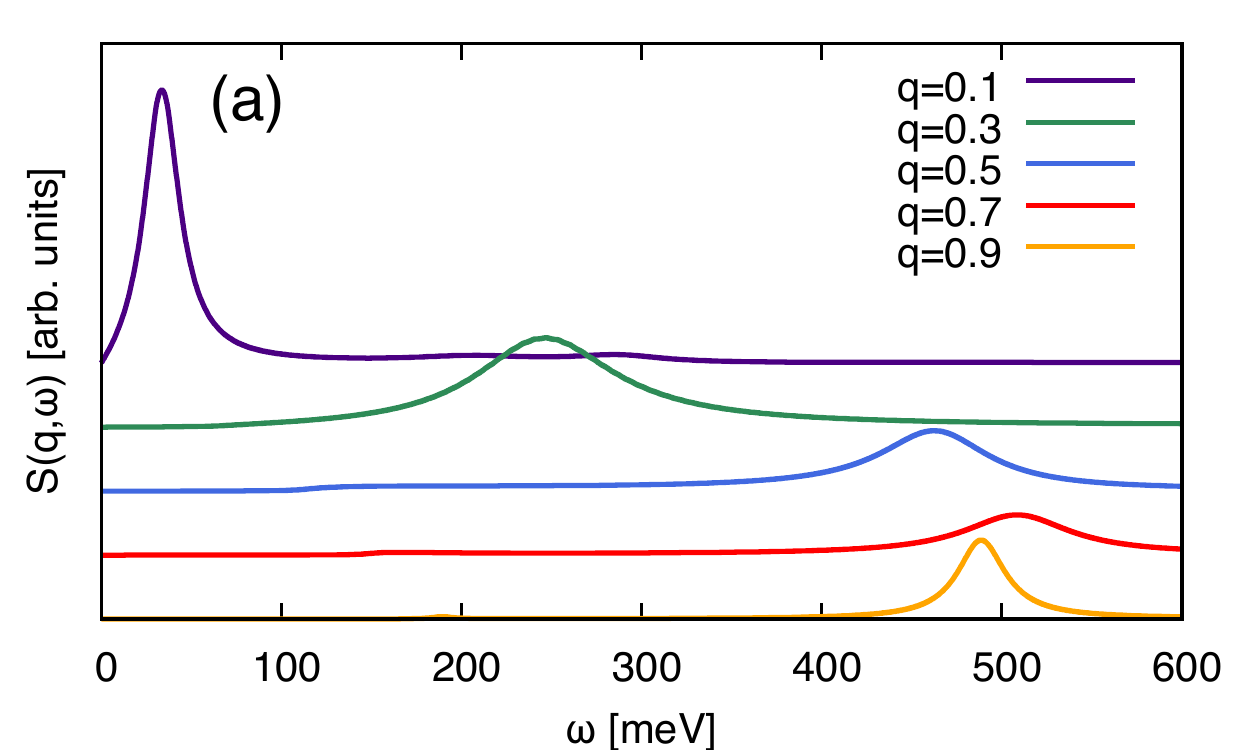}
\includegraphics[width=0.98\columnwidth]{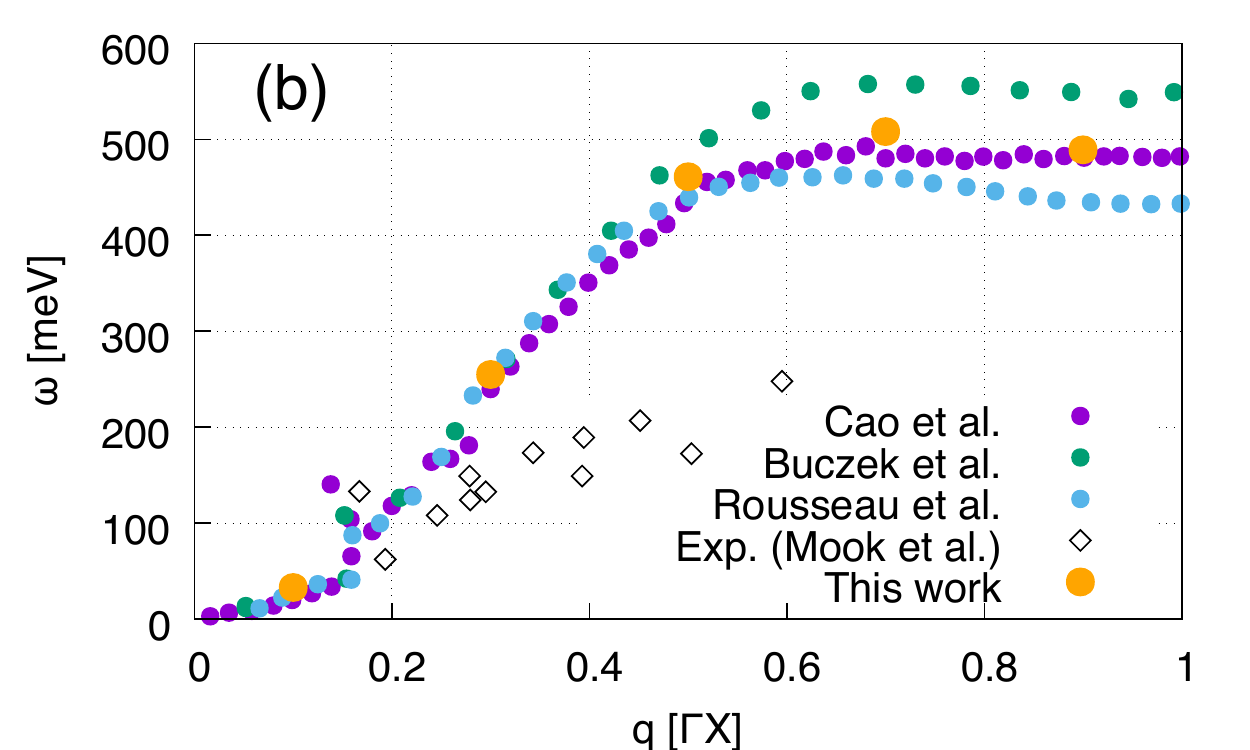}
\caption{Magnon spectra and dispersion in Ni along the $\Gamma$-X direction in the Brillouin zone. (a)~Excitation spectrum $S(\mathbf{q},\omega)$ at several values of the transferred momentum $\mathbf{q} = \frac{2\pi}{a} (q,0,0)$. (b)~Comparison of the magnon dispersion as obtained in this work, in other theoretical works (Cao et al.~\cite{Cao:2018}, Buczek et al.~\cite{Buczek:2011b}, Rousseau et al.~\cite{Rousseau:2012}), and as measured in the INSS experiment at 300~K (Mook et al.~\cite{Mook:1985}). Each point in (b) represents the position of the maximum of $S(\mathbf{q},\omega)$ in (a).}
\label{fig:Ni_magnon}
\end{figure}

In this section we show the magnon dispersions for bulk Fe and Ni, which are obtained after the convergence tests with respect to the number of Lanczos iterations and using the extrapolation technique as described in the previous section. The magnon spectra at various values of the transferred momenta $\mathbf{q}$ along the [110] direction for Fe and along the [100] direction for Ni are shown in Figs.~\ref{fig:Fe_magnon} and \ref{fig:Ni_magnon}, respectively.
We note that only one Lanczos chain is needed for each value of the transferred momentum $\mathbf{q}$, since in ferromagnetic collinear structures only the response to the external magnetic field perpendicular to both $\mathbf{q}$ and the ground-state magnetization contributes to the excitation spectrum (as long as $\mathbf{q}$ and the ground-state magnetization are non-parallel).

In agreement with previous calculations and experiments, the magnon peak of both Fe and Ni is sharp at small values of the transferred momenta, whereas it becomes damped at larger values of the transferred momenta [see Figs.~\ref{fig:Fe_magnon}~(a) and \ref{fig:Ni_magnon}~(a)]. As it is well known the damping of the magnon occurs when it enters in the Stoner continuum, which leads to the fact that the energy of the magnon is transferred to the creation of the electron-hole pairs. In Figs.~\ref{fig:Fe_magnon}~(b) and \ref{fig:Ni_magnon}~(b) we show the magnon dispersion in Fe and Ni, respectively, as obtained in this work using the Liouville-Lanczos approach, in other TDDFT studies~\cite{Rousseau:2012, Buczek:2011b, Cao:2018}, and in the INSS experiments~\cite{Loong:1984, Mook:1985}. It can be seen that our calculations are in good agreement with other TDDFT studies both for Fe and Ni, though there are some variations between all the theoretical results which may be due to differences in the details of the implementation and in the computational parameters used. In particular, for Ni there are variations in the magnon dispersion close to the edge of the Brillouin zone: in our calculations, and in agreement with Refs.~\cite{Buczek:2011b, Rousseau:2012}, we find a small decrease in the magnon energy while in Ref.~\cite{Cao:2018} a plateau-like magnon dispersion was observed - these small discrepancies might be attributed to the differences in the $\mathbf{k}$ point sampling of the Brillouin zone and the smearing technique used (as \emph{e.g.} in Ref.~\cite{Cao:2018} a frequency-dependent smearing was used).

For Fe our magnon dispersion is in very good agreement with the experimental data of Ref.~\cite{Loong:1984}. However, for Ni the agreement between our calculations (as well as all other TDDFT studies) and the experiments is good only at small values of the transferred momenta $\mathbf{q}$, while at larger values of $\mathbf{q}$ the theoretical magnon energies overestimate the experimental ones due to the overestimation of the exchange splitting when using local spin density approximation~\cite{Karlsson:2000, Sasioglu:2010, Eich:2018}. More advanced {\it ab initio} approaches for a more accurate treatment of the exchange splitting are therefore required, in order to overcome this drawback. Finally, it is worth noting that for Ni, when having a more dense $\mathbf{q}$ point sampling of the magnon dispersion and when using smaller values of the broadening parameter for the magnon spectra, there are evidences of a presence of two magnon branches - acoustic and optical~\cite{Savrasov:1998, Karlsson:2000, Sasioglu:2010}, which though we do not attempt to resolve in our calculations.

Some attention has been paid in the literature to the violation of the Goldstone theorem that some authors find when computing the magnon dispersion in the long-wavelength limit~\cite{Lounis:2011, Buczek:2011b, Rousseau:2012, Muller:2016}. This violation is generally due to different numerical and/or physical approximations used to deal with the ground and excited states (such as \emph{e.g.} different grids of $\mathbf{k}$ points or truncation of the susceptibilities when solving the Dyson equation or inconsistent response functions). Our approach, as well as the one based on the Sternheimer equation~\cite{Savrasov:1998, Cao:2018}, derives directly from the linearization of the time-dependent Kohn-Sham equations [Eqs.~(\ref{eq:KS_eq_resonant_projected_q}-\ref{eq:KS_eq_antiresonant_projected_q})], which, in the absence of spin-orbit coupling, are invariant with respect to spin rotations, thus enforcing a zero magnon frequency in the long-wavelength limit. Our numerical tests for Fe and Ni at $|\mathbf{q}| = 0.01\times \frac{2\pi}{a}$ \, using the Liouville-Lanczos approach are consistent with a vanishing magnon frequency. 

\section{Conclusions}
\label{sec:conclusions}

We believe that the Liouville-Lanczos approach introduced in this paper will offer substantial advantages with respect to existing techniques to model spin-wave excitations in complex magnetic systems, both in the bulk and in reduced dimensionality. This approach presents conceptual similarities with the one based on the Sternheimer equation presented in Refs.~\cite{Savrasov:1998,Cao:2018}. In a sense, the present method amounts to solving the Sternheimer equation in a Krylov subspace using a basis that allows one to make the bulk of the numerical work independent of the frequency at which the equation is solved. The continued-fraction representation of the magnetic susceptibility resulting from the inversion of the tridiagonal matrix in Eq.~\eqref{eq:resolvent_g} can be thus seen as a Pad\'e interpolation of the results obtained by solving the Sternheimer equation at different frequencies. It is possible that pursuing these analogies will result in valuable computational savings.

The Liouville-Lanczos approach has been developed and implemented in a fully general spinor formalism, so that it is ready-to-use for systems in which spin-orbit coupling effects are important. The main feature of the new method is that a single Lanczos chain is needed to obtain the full magnetic spectrum (over a broad frequency range) for each magnon wave-vector and polarization. In addition to the already considerable numerical and conceptual advantages, we think that there is ample room for further improvements, including an improved sampling of the response over the Brillouin zone (by both interpolating the response at different electron wave-vectors and by leveraging crystal symmetry), and by improving the convergence of the Lanczos recursion by optimizing the representation of the response orbitals and reducing the condition number of the Liouvillian. Work along all of these lines is in progress.

\section*{ACKNOWLEDGMENTS}

We thank Andrea~Dal~Corso, Stefano~de~Gironcoli, and Paolo~Giannozzi for valuable discussions, and Carlo Cavazzoni at CINECA for technical support. This work was partially funded by the EU through the \textsc{MaX} Centre of Excellence for supercomputing applications (Project No.~676598), and the NCCR MARVEL. Computer time was partially provided by CINECA through grant No.~IsC38. TG acknowledges partial support by the Cluster of Excellence MATISSE programme led by Sorbonne Universit\'es (reference ANR-11-IDEX-0004-02).

\appendix

\section{Generalized spin-charge susceptibility}
\label{sec:general_pert}

Similarly to Refs.~\cite{Rousseau:2012, dosSantosDias:2015, Cao:2018}, our formalism presented in the main text can be straightforwardly generalized to a more general perturbation, which contains the scalar perturbing potential $\tilde{v}^{\prime}_{\mathrm{ext},\mathbf{q}}(\mathbf{r}, \omega)$ and the external magnetic field $\tilde{\boldsymbol b}'_{\mathrm{ext},\mathbf{q}}(\mathbf{r}, \omega)$ [see Eqs.~\eqref{eq:V_ext_q-decomposition} and \eqref{eq:V_ext_q}]:  
\begin{equation}
\tilde{V}^{\prime}_{\mathrm{ext},\mathbf{q}}(\mathbf{r}, \omega) =
\sigma^\circ \, \tilde{v}^{\prime}_{\mathrm{ext},\mathbf{q}}(\mathbf{r}, \omega)
- \mu_\mathrm{B} \, {\boldsymbol \sigma} \cdot
\tilde{\boldsymbol b}'_{\mathrm{ext},\mathbf{q}}(\mathbf{r}, \omega) \,.
\label{eq:V_ext_q_general}
\end{equation}
In this case, all the formalities discussed in Sec.~\ref{sec:general_formulation} remain valid given that Eq.~\eqref{eq:V_ext_q_general} is used instead of Eq.~\eqref{eq:V_ext_q}.

By defining a generic 4-component operator $\hat{\boldsymbol A}_\mathbf{q} = \{ \hat{n}_\mathbf{q}, \hat{\boldsymbol m}_\mathbf{q} \}$, we can compute the expectation value of its response by generalizing Eq.~\eqref{eq:m_expectation_value} to:
\begin{eqnarray}
\bigl\langle \hat{{\boldsymbol A}'}_\mathbf{q} \bigr\rangle_\omega & = &
\mathrm{Tr}[\hat{\boldsymbol A}_{\mathbf{q}}^\dagger \,
\hat{\rho}_{\mathbf{q}}'(\omega)] \nonumber \\
& = & \left( \hat{\boldsymbol A}_{\mathbf{q}}, (\omega - \hat{\mathcal{L}}_\mathbf{q})^{-1} \cdot
[\hat{\tilde{V}}^{\prime}_\mathrm{ext, \mathbf{q}}(\omega), \hat{\rho}^\circ] \right) \,.
\label{eq:A_expectation_value}
\end{eqnarray}
Using arguments similar to those in Sec.~\ref{sec:quantum_Liouville_eq}, we can define a $4 \times 4$ generalized susceptibility:
\begin{equation}
{\boldsymbol \chi}_A(\mathbf{q}, \mathbf{q}; \omega) =
\left( \hat{\boldsymbol A}_{\mathbf{q}}, (\omega - \hat{\mathcal{L}_\mathbf{q}})^{-1} \cdot
[\hat{\boldsymbol A}_{\mathbf{q}}, \hat{\rho}^\circ] \right) \,,
\label{eq:spin-charge-density_susceptibility}
\end{equation}
which contains charge-charge, spin-spin, charge-spin, and spin-charge responses. Typically, charge-charge (plasmons) and spin-spin (magnons) responses do not overlap in energies, because the former appear at several eV or tens of eV, while the latter appear at several tens or hundreds of meV. However, if plasmons' and magnons' energies start overlapping then a generalized description presented here becomes important.

The batch representation of Sec.~\ref{sec:batch} can be straightforwardly generalized to this case and subsequently used with the Lanczos algorithm of Sec.~\ref{sec:Lanczos_algorithm}.

\section{Numerical treatment of metals}
\label{sec:metals}

In this Appendix we present a generalization of the Liouville-Lanczos approach within TDDFpT to magnetic excitations in metals.

When considering metals, special care must be taken of sums
over $\mathbf{k}$ points around the Fermi surface in the Brillouin zone.
In practice, various smearing techniques are used in order to speed up the convergence
of such sums~\cite{Fu:1983,Methfessel:1989,Marzari:1999}. This implies adding extra
complexity in the TDDFpT formalism, which we discuss in the following. 

Let us start from the ground-state charge and magnetization densities,
Eqs.~\eqref{eq:charge_dens_0} and \eqref{eq:spin_dens_0}. Using smearing techniques implies
replacing the zero-temperature occupations $f_{n,\mathbf{k}}$ with the smearing step-like functions $\tilde{\theta}_{n,\mathbf{k}}$ which mimic some small finite temperature controlled 
by $\eta$ (called a broadening parameter). Therefore, using Eq.~\eqref{eq:Psi_GS_Bloch}, 
for metals we have:
\begin{equation}
n^\circ(\mathbf{r}) = \frac{1}{N_\mathbf{k}}
\sum_{n,\mathbf{k}} \tilde{\theta}_{n,\mathbf{k}} \,
U_{n, \mathbf{k}}^{\circ \dagger}(\mathbf{r}) \, U_{n, \mathbf{k}}^{\circ}(\mathbf{r}) \,,
\label{eq:charge_dens_0_met}
\end{equation}
\begin{equation}
{\boldsymbol m}^\circ(\mathbf{r}) =
\frac{\mu_\mathrm{B}}{N_\mathbf{k}} \sum_{n,\mathbf{k}}
\tilde{\theta}_{n,\mathbf{k}} \,
U_{n, \mathbf{k}}^{\circ \dagger}(\mathbf{r}) \,
{\boldsymbol \sigma} \, U_{n, \mathbf{k}}^{\circ}(\mathbf{r}) \,,
\label{eq:spin_dens_0_met}
\end{equation}
where  $\tilde{\theta}_{n,\mathbf{k}} \equiv \tilde{\theta}[(\varepsilon_F^\circ - \varepsilon_{n,\mathbf{k}}^\circ)/\eta]$, with $\tilde{\theta}$ being the
smooth function~\cite{Fu:1983, Methfessel:1989, Marzari:1999} which approximates the step-like function, $\varepsilon_F^\circ$ is the Fermi energy,
and the summation with $n$ runs over all fully occupied states plus
a small number of partially occupied states (\emph{e.g.} 20\% of the number of fully occupied states).
Similar replacement of $f_{n,\mathbf{k}}$ by $\tilde{\theta}_{n,\mathbf{k}}$ must be done 
in the expressions for the response charge and magnetization densities [see Eqs.~\eqref{eq:charge_dens_resp_q} and \eqref{eq:spin_dens_resp_q}]. However, it is convenient to redefine the response KS spinor wave functions appearing in these equations in such a way that the smearing functions $\tilde{\theta}_{n,\mathbf{k}}$ are not present explicitly in Eqs.~\eqref{eq:charge_dens_resp_q} and \eqref{eq:spin_dens_resp_q}. This is the same idea as 
in the static DFpT for metals~\cite{deGironcoli:1995, Baroni:2001} and
its generalization to the frequency domain~\cite{Timrov:2013}. By doing so, we can rewrite the response charge and magnetization densities for metals as: 
\begin{widetext}
\begin{align}
    \tilde{n}'_\mathbf{q}(\mathbf{r},\omega) & =
\frac{1}{N_\mathbf{k}} \sum_{n, \mathbf{k}} \biggl[ U^{\circ \dagger}_{n,\mathbf{k}}(\mathbf{r}) \, \tilde{U}'_{n,\mathbf{k+q}}(\mathbf{r},\omega) + \left(\hat{\rm T}U^\circ_{n,-\mathbf{k}}(\mathbf{r}) \right)^{\dag} \hat{\mathrm{T}}\tilde{U}'_{n,-\mathbf{k}-\mathbf{q}}(\mathbf{r},-\omega) \biggr] \,,
\label{eq:charge_dens_resp_q_metal} \\
    \tilde{\boldsymbol m}'_\mathbf{q}(\mathbf{r},\omega) & =
\frac{\mu_\mathrm{B}}{N_\mathbf{k}} \sum_{n, \mathbf{k}}
\biggl[ U^{\circ \dagger}_{n,\mathbf{k}}(\mathbf{r}) \, {\boldsymbol \sigma} \, \tilde{U}'_{n,\mathbf{k+q}}(\mathbf{r},\omega) - \left(\hat{\rm T}U^\circ_{n,-\mathbf{k}}(\mathbf{r}) \right)^{\dag} \, {\boldsymbol \sigma} \, \hat{\mathrm{T}}\tilde{U}'_{n,-\mathbf{k}-\mathbf{q}}(\mathbf{r},-\omega) \biggr] \,,
\label{eq:spin_dens_resp_q_metal}
\end{align}
where now the response KS spinor wave functions $\tilde{U}'_{n,\mathbf{k+q}}(\mathbf{r},\omega)$ and $\hat{\mathrm{T}} \tilde{U}'_{n,-\mathbf{k}-\mathbf{q}}(\mathbf{r},-\omega)$ satisfy
the linearized KS equations for metals:
\begin{align}
\bigl( \hat{H}^{\circ}_\mathbf{k+q} - \varepsilon^\circ_{n,\mathbf{k}} - \omega \bigr) \,
\tilde{U}'_{n,\mathbf{k+q}}(\mathbf{r},\omega) + \, \hat{P}_{n,\mathbf{k},\mathbf{k+q}} \,
\hat{\tilde{V}}^{\prime}_{\mathrm{HXC},\mathbf{q}}(\omega) \,
U^\circ_{n,\mathbf{k}}(\mathbf{r})
& = - \hat{P}_{n,\mathbf{k},\mathbf{k+q}} \,
\hat{\tilde{V}}^{\prime}_{\mathrm{ext},\mathbf{q}}(\omega)
\, U^\circ_{n,\mathbf{k}}(\mathbf{r}) \,,
\label{eq:KS_eq_resonant_projected_q_metals} \\
\bigl( \hat{H}^{\circ +}_\mathbf{k+q} - \varepsilon^\circ_{n,-\mathbf{k}} + \omega \bigr) \,
\hat{\mathrm{T}} \tilde{U}'_{n,-\mathbf{k}-\mathbf{q}}(\mathbf{r},-\omega) +
\, \hat{P}^{+}_{n,\mathbf{k},\mathbf{k+q}} \,
\hat{\tilde{V}}^{\prime +}_{\mathrm{HXC},\mathbf{q}}(\omega) \,
\hat{\mathrm{T}} U^\circ_{n,-\mathbf{k}}(\mathbf{r})
&= - \hat{P}^{+}_{n,\mathbf{k},\mathbf{k+q}} \,
\hat{\tilde{V}}^{\prime +}_{\mathrm{ext},\mathbf{q}}(\omega) \,
\hat{\mathrm{T}} U^\circ_{n,-\mathbf{k}}(\mathbf{r}) \,.
\label{eq:KS_eq_antiresonant_projected_q_metals}
\end{align}
Note that, with respect to Eqs.~\eqref{eq:KS_eq_resonant_projected_q} --
\eqref{eq:KS_eq_antiresonant_projected_q}, here we have introduced two new
operators, $\hat{P}_{n,\mathbf{k},\mathbf{k+q}}$ and
$\hat{P}^{+}_{n,\mathbf{k},\mathbf{k+q}}$, which in the coordiante
representation are defined as:
\begin{align}
    P_{n,\mathbf{k},\mathbf{k+q}}(\mathbf{r},\mathbf{r}') & =
\tilde{\theta}_{n,\mathbf{k}} \, \delta(\mathbf{r}-\mathbf{r}')
- \sum_{m} \beta_{n,\mathbf{k}; m,\mathbf{k+q}} \, U^\circ_{m,\mathbf{k+q}}(\mathbf{r}) \,
U^{\circ \dagger}_{m,\mathbf{k+q}}(\mathbf{r}') \,, \label{eq:P_minus_q} \\
P^{+}_{n,\mathbf{k},\mathbf{k+q}}(\mathbf{r},\mathbf{r}') & =
\hat{\rm T} \, P_{n,-\mathbf{k},-\mathbf{k}-\mathbf{q}}(\mathbf{r},\mathbf{r}')
\, \hat{\rm T}^{-1} \nonumber \\
& = \tilde{\theta}_{n,-\mathbf{k}} \, \delta(\mathbf{r}-\mathbf{r}') -
\sum_{m} \beta_{n,-\mathbf{k}; m,-\mathbf{k}-\mathbf{q}}
\left( \hat{\mathrm{T}} U^\circ_{n,-\mathbf{k}-\mathbf{q}}(\mathbf{r}) \right)
\left( \hat{\mathrm{T}} U^{\circ}_{n,-\mathbf{k}-\mathbf{q}}(\mathbf{r}') \right)^\dagger \,,
 \label{eq:P_plus_q}
\end{align}
where
\begin{equation}
\beta_{n,\mathbf{k};m,\mathbf{k+q}} =
\tilde{\theta}_{n,\mathbf{k}} \, \theta_{n,\mathbf{k};m,\mathbf{k+q}} +
\, \tilde{\theta}_{m,\mathbf{k+q}} \, \theta_{m,\mathbf{k+q};n,\mathbf{k}} \,.
\label{eq:beta_def_metals}
\end{equation}
\end{widetext}
In Eq.~\eqref{eq:beta_def_metals}, $\theta_{n,\mathbf{k}; m, \mathbf{k+q}} \equiv \theta[(\varepsilon^\circ_{n,\mathbf{k}} - \varepsilon^\circ_{m, \mathbf{k+q}})/\eta]$
is the rescaled complementary error function with the opposite sign of the argument,
i.e. $\theta(\varepsilon) \equiv \mathrm{erfc}(-\varepsilon)/2$ \cite{Note:notations_theta}.
In practice, the summations in Eqs.~\eqref{eq:P_minus_q} and \eqref{eq:P_plus_q} run over all the states up to $\varepsilon_F$ plus the partially occupied states with energy in the range from $\varepsilon_F$ to
$\varepsilon_F + 3\eta$~\cite{deGironcoli:1995}. Thus, only a few partially occupied states
above the Fermi level are needed, while the whole manifold of fully empty states needs not to be computed,
which is one of the main bottlenecks of the state-of-the-art methods as those based on the
solution of the Dyson equation~\cite{Buczek:2011b, dosSantosDias:2015, Wysocki:2017}.


\begin{thebibliography}{62}%
\makeatletter
\providecommand \@ifxundefined [1]{%
 \@ifx{#1\undefined}
}%
\providecommand \@ifnum [1]{%
 \ifnum #1\expandafter \@firstoftwo
 \else \expandafter \@secondoftwo
 \fi
}%
\providecommand \@ifx [1]{%
 \ifx #1\expandafter \@firstoftwo
 \else \expandafter \@secondoftwo
 \fi
}%
\providecommand \natexlab [1]{#1}%
\providecommand \enquote  [1]{``#1''}%
\providecommand \bibnamefont  [1]{#1}%
\providecommand \bibfnamefont [1]{#1}%
\providecommand \citenamefont [1]{#1}%
\providecommand \href@noop [0]{\@secondoftwo}%
\providecommand \href [0]{\begingroup \@sanitize@url \@href}%
\providecommand \@href[1]{\@@startlink{#1}\@@href}%
\providecommand \@@href[1]{\endgroup#1\@@endlink}%
\providecommand \@sanitize@url [0]{\catcode `\\12\catcode `\$12\catcode
  `\&12\catcode `\#12\catcode `\^12\catcode `\_12\catcode `\%12\relax}%
\providecommand \@@startlink[1]{}%
\providecommand \@@endlink[0]{}%
\providecommand \url  [0]{\begingroup\@sanitize@url \@url }%
\providecommand \@url [1]{\endgroup\@href {#1}{\urlprefix }}%
\providecommand \urlprefix  [0]{URL }%
\providecommand \Eprint [0]{\href }%
\providecommand \doibase [0]{http://dx.doi.org/}%
\providecommand \selectlanguage [0]{\@gobble}%
\providecommand \bibinfo  [0]{\@secondoftwo}%
\providecommand \bibfield  [0]{\@secondoftwo}%
\providecommand \translation [1]{[#1]}%
\providecommand \BibitemOpen [0]{}%
\providecommand \bibitemStop [0]{}%
\providecommand \bibitemNoStop [0]{.\EOS\space}%
\providecommand \EOS [0]{\spacefactor3000\relax}%
\providecommand \BibitemShut  [1]{\csname bibitem#1\endcsname}%
\let\auto@bib@innerbib\@empty
\bibitem [{\citenamefont {Moriya}(1985)}]{Moriya:1985}%
  \BibitemOpen
  \bibfield  {author} {\bibinfo {author} {\bibfnamefont {T.}~\bibnamefont
  {Moriya}},\ }\href@noop {} {\emph {\bibinfo {title} {Spin Fluctuations in
  Itinerant Electron Magnetism}}}\ (\bibinfo  {publisher} {Springer-Verlag},\
  \bibinfo {address} {Berlin},\ \bibinfo {year} {1985})\BibitemShut {NoStop}%
\bibitem [{\citenamefont {Zakeri}(2014)}]{Zakeri:2014}%
  \BibitemOpen
  \bibfield  {author} {\bibinfo {author} {\bibfnamefont {K.}~\bibnamefont
  {Zakeri}},\ }\href@noop {} {\bibfield  {journal} {\bibinfo  {journal}
  {Physics Reports}\ }\textbf {\bibinfo {volume} {545}},\ \bibinfo {pages} {47}
  (\bibinfo {year} {2014})}\BibitemShut {NoStop}%
\bibitem [{\citenamefont {Mook}\ and\ \citenamefont
  {Nicklow}(1973)}]{Mook:1973}%
  \BibitemOpen
  \bibfield  {author} {\bibinfo {author} {\bibfnamefont {H.}~\bibnamefont
  {Mook}}\ and\ \bibinfo {author} {\bibfnamefont {R.}~\bibnamefont {Nicklow}},\
  }\href@noop {} {\bibfield  {journal} {\bibinfo  {journal} {Phys. Rev. B}\
  }\textbf {\bibinfo {volume} {7}},\ \bibinfo {pages} {336} (\bibinfo {year}
  {1973})}\BibitemShut {NoStop}%
\bibitem [{\citenamefont {Qin}\ \emph {et~al.}(2015)\citenamefont {Qin},
  \citenamefont {Zakeri}, \citenamefont {Ernst}, \citenamefont {Sandratskii},
  \citenamefont {Buczek}, \citenamefont {Marmodoro}, \citenamefont {Chuang},
  \citenamefont {Zhang},\ and\ \citenamefont {Kirschner}}]{Qin:2015}%
  \BibitemOpen
  \bibfield  {author} {\bibinfo {author} {\bibfnamefont {H.}~\bibnamefont
  {Qin}}, \bibinfo {author} {\bibfnamefont {K.}~\bibnamefont {Zakeri}},
  \bibinfo {author} {\bibfnamefont {A.}~\bibnamefont {Ernst}}, \bibinfo
  {author} {\bibfnamefont {L.}~\bibnamefont {Sandratskii}}, \bibinfo {author}
  {\bibfnamefont {P.}~\bibnamefont {Buczek}}, \bibinfo {author} {\bibfnamefont
  {A.}~\bibnamefont {Marmodoro}}, \bibinfo {author} {\bibfnamefont {T.-H.}\
  \bibnamefont {Chuang}}, \bibinfo {author} {\bibfnamefont {Y.}~\bibnamefont
  {Zhang}}, \ and\ \bibinfo {author} {\bibfnamefont {J.}~\bibnamefont
  {Kirschner}},\ }\href@noop {} {\bibfield  {journal} {\bibinfo  {journal}
  {Nat. Commun.}\ }\textbf {\bibinfo {volume} {6}},\ \bibinfo {pages} {6126}
  (\bibinfo {year} {2015})}\BibitemShut {NoStop}%
\bibitem [{\citenamefont {Hirjibehedin}\ \emph {et~al.}(2006)\citenamefont
  {Hirjibehedin}, \citenamefont {Lutz},\ and\ \citenamefont
  {Heinrich}}]{Hirjibehedin:2006}%
  \BibitemOpen
  \bibfield  {author} {\bibinfo {author} {\bibfnamefont {C.}~\bibnamefont
  {Hirjibehedin}}, \bibinfo {author} {\bibfnamefont {J.}~\bibnamefont {Lutz}},
  \ and\ \bibinfo {author} {\bibfnamefont {A.}~\bibnamefont {Heinrich}},\
  }\href@noop {} {\bibfield  {journal} {\bibinfo  {journal} {Science}\ }\textbf
  {\bibinfo {volume} {312}},\ \bibinfo {pages} {1021} (\bibinfo {year}
  {2006})}\BibitemShut {NoStop}%
\bibitem [{\citenamefont {Landau}(1946)}]{Landau:1946}%
  \BibitemOpen
  \bibfield  {author} {\bibinfo {author} {\bibfnamefont {L.}~\bibnamefont
  {Landau}},\ }\href@noop {} {\bibfield  {journal} {\bibinfo  {journal} {J.
  Phys. USSR}\ }\textbf {\bibinfo {volume} {10}},\ \bibinfo {pages} {25}
  (\bibinfo {year} {1946})}\BibitemShut {NoStop}%
\bibitem [{\citenamefont {Fetter}\ and\ \citenamefont
  {Walecka}(1971)}]{Fetter:1971}%
  \BibitemOpen
  \bibfield  {author} {\bibinfo {author} {\bibfnamefont {A.}~\bibnamefont
  {Fetter}}\ and\ \bibinfo {author} {\bibfnamefont {J.}~\bibnamefont
  {Walecka}},\ }\href@noop {} {\emph {\bibinfo {title} {{Q}uantum {T}heory of
  {M}any-{P}articles {S}ystems}}}\ (\bibinfo  {publisher} {{I}nternational
  {S}eries in {P}ure and {A}pplied {P}hysics},\ \bibinfo {address}
  {McGraw-Hill, New York},\ \bibinfo {year} {1971})\BibitemShut {NoStop}%
\bibitem [{\citenamefont {Costa}\ \emph {et~al.}(2010)\citenamefont {Costa},
  \citenamefont {Muniz}, \citenamefont {Lounis}, \citenamefont {Klautau},\ and\
  \citenamefont {Mills}}]{Costa:2010}%
  \BibitemOpen
  \bibfield  {author} {\bibinfo {author} {\bibfnamefont {A.}~\bibnamefont
  {Costa}}, \bibinfo {author} {\bibfnamefont {R.}~\bibnamefont {Muniz}},
  \bibinfo {author} {\bibfnamefont {S.}~\bibnamefont {Lounis}}, \bibinfo
  {author} {\bibfnamefont {A.}~\bibnamefont {Klautau}}, \ and\ \bibinfo
  {author} {\bibfnamefont {D.}~\bibnamefont {Mills}},\ }\href@noop {}
  {\bibfield  {journal} {\bibinfo  {journal} {Phys. Rev. B}\ }\textbf {\bibinfo
  {volume} {82}},\ \bibinfo {pages} {014428} (\bibinfo {year}
  {2010})}\BibitemShut {NoStop}%
\bibitem [{\citenamefont {Bergman}\ \emph {et~al.}(2010)\citenamefont
  {Bergman}, \citenamefont {Taroni}, \citenamefont {Bergqvist}, \citenamefont
  {Hellsvik}, \citenamefont {Hj\"orvarsson},\ and\ \citenamefont
  {Eriksson}}]{Bergman:2010}%
  \BibitemOpen
  \bibfield  {author} {\bibinfo {author} {\bibfnamefont {A.}~\bibnamefont
  {Bergman}}, \bibinfo {author} {\bibfnamefont {A.}~\bibnamefont {Taroni}},
  \bibinfo {author} {\bibfnamefont {L.}~\bibnamefont {Bergqvist}}, \bibinfo
  {author} {\bibfnamefont {J.}~\bibnamefont {Hellsvik}}, \bibinfo {author}
  {\bibfnamefont {B.}~\bibnamefont {Hj\"orvarsson}}, \ and\ \bibinfo {author}
  {\bibfnamefont {O.}~\bibnamefont {Eriksson}},\ }\href@noop {} {\bibfield
  {journal} {\bibinfo  {journal} {Phys. Rev. B}\ }\textbf {\bibinfo {volume}
  {81}},\ \bibinfo {pages} {144416} (\bibinfo {year} {2010})}\BibitemShut
  {NoStop}%
\bibitem [{\citenamefont {Zakeri}\ \emph {et~al.}(2012)\citenamefont {Zakeri},
  \citenamefont {Zhang}, \citenamefont {Chuang},\ and\ \citenamefont
  {Kirschner}}]{Zakeri:2012}%
  \BibitemOpen
  \bibfield  {author} {\bibinfo {author} {\bibfnamefont {K.}~\bibnamefont
  {Zakeri}}, \bibinfo {author} {\bibfnamefont {Y.}~\bibnamefont {Zhang}},
  \bibinfo {author} {\bibfnamefont {T.-H.}\ \bibnamefont {Chuang}}, \ and\
  \bibinfo {author} {\bibfnamefont {J.}~\bibnamefont {Kirschner}},\ }\href@noop
  {} {\bibfield  {journal} {\bibinfo  {journal} {Phys. Rev. Lett.}\ }\textbf
  {\bibinfo {volume} {108}},\ \bibinfo {pages} {197205} (\bibinfo {year}
  {2012})}\BibitemShut {NoStop}%
\bibitem [{\citenamefont {Zakeri}(2017)}]{Zakeri:2017}%
  \BibitemOpen
  \bibfield  {author} {\bibinfo {author} {\bibfnamefont {K.}~\bibnamefont
  {Zakeri}},\ }\href@noop {} {\bibfield  {journal} {\bibinfo  {journal} {J.
  Phys.: Condens. Matter}\ }\textbf {\bibinfo {volume} {29}},\ \bibinfo {pages}
  {013001} (\bibinfo {year} {2017})}\BibitemShut {NoStop}%
\bibitem [{\citenamefont {Niu}\ and\ \citenamefont
  {Kleinman}(1998)}]{Niu:1998}%
  \BibitemOpen
  \bibfield  {author} {\bibinfo {author} {\bibfnamefont {Q.}~\bibnamefont
  {Niu}}\ and\ \bibinfo {author} {\bibfnamefont {L.}~\bibnamefont {Kleinman}},\
  }\href {\doibase 10.1103/physrevlett.80.2205} {\bibfield  {journal} {\bibinfo
   {journal} {Phys. Rev. Lett.}\ }\textbf {\bibinfo {volume} {80}},\ \bibinfo
  {pages} {2205} (\bibinfo {year} {1998})}\BibitemShut {NoStop}%
\bibitem [{\citenamefont {Gebauer}\ and\ \citenamefont
  {Baroni}(2000)}]{Gebauer:2000}%
  \BibitemOpen
  \bibfield  {author} {\bibinfo {author} {\bibfnamefont {R.}~\bibnamefont
  {Gebauer}}\ and\ \bibinfo {author} {\bibfnamefont {S.}~\bibnamefont
  {Baroni}},\ }\href@noop {} {\bibfield  {journal} {\bibinfo  {journal} {Phys.
  Rev. B}\ }\textbf {\bibinfo {volume} {61}},\ \bibinfo {pages} {R6459}
  (\bibinfo {year} {2000})}\BibitemShut {NoStop}%
\bibitem [{\citenamefont {Savrasov}(1998)}]{Savrasov:1998}%
  \BibitemOpen
  \bibfield  {author} {\bibinfo {author} {\bibfnamefont {S.}~\bibnamefont
  {Savrasov}},\ }\href@noop {} {\bibfield  {journal} {\bibinfo  {journal}
  {Phys. Rev. Lett.}\ }\textbf {\bibinfo {volume} {81}},\ \bibinfo {pages}
  {2570} (\bibinfo {year} {1998})}\BibitemShut {NoStop}%
\bibitem [{\citenamefont {Lounis}\ \emph {et~al.}(2011)\citenamefont {Lounis},
  \citenamefont {Costa}, \citenamefont {Muniz},\ and\ \citenamefont
  {Mills}}]{Lounis:2011}%
  \BibitemOpen
  \bibfield  {author} {\bibinfo {author} {\bibfnamefont {S.}~\bibnamefont
  {Lounis}}, \bibinfo {author} {\bibfnamefont {A.}~\bibnamefont {Costa}},
  \bibinfo {author} {\bibfnamefont {R.}~\bibnamefont {Muniz}}, \ and\ \bibinfo
  {author} {\bibfnamefont {D.}~\bibnamefont {Mills}},\ }\href@noop {}
  {\bibfield  {journal} {\bibinfo  {journal} {Phys. Rev. B}\ }\textbf {\bibinfo
  {volume} {83}},\ \bibinfo {pages} {035109} (\bibinfo {year}
  {2011})}\BibitemShut {NoStop}%
\bibitem [{\citenamefont {Buczek}\ \emph {et~al.}(2011)\citenamefont {Buczek},
  \citenamefont {Ernst},\ and\ \citenamefont {Sandratskii}}]{Buczek:2011b}%
  \BibitemOpen
  \bibfield  {author} {\bibinfo {author} {\bibfnamefont {P.}~\bibnamefont
  {Buczek}}, \bibinfo {author} {\bibfnamefont {A.}~\bibnamefont {Ernst}}, \
  and\ \bibinfo {author} {\bibfnamefont {L.}~\bibnamefont {Sandratskii}},\
  }\href@noop {} {\bibfield  {journal} {\bibinfo  {journal} {Phys. Rev. B}\
  }\textbf {\bibinfo {volume} {84}},\ \bibinfo {pages} {174418} (\bibinfo
  {year} {2011})}\BibitemShut {NoStop}%
\bibitem [{\citenamefont {Rousseau}\ \emph {et~al.}(2012)\citenamefont
  {Rousseau}, \citenamefont {Eiguren},\ and\ \citenamefont
  {Bergara}}]{Rousseau:2012}%
  \BibitemOpen
  \bibfield  {author} {\bibinfo {author} {\bibfnamefont {B.}~\bibnamefont
  {Rousseau}}, \bibinfo {author} {\bibfnamefont {A.}~\bibnamefont {Eiguren}}, \
  and\ \bibinfo {author} {\bibfnamefont {A.}~\bibnamefont {Bergara}},\
  }\href@noop {} {\bibfield  {journal} {\bibinfo  {journal} {Phys. Rev. B}\
  }\textbf {\bibinfo {volume} {85}},\ \bibinfo {pages} {054305} (\bibinfo
  {year} {2012})}\BibitemShut {NoStop}%
\bibitem [{\citenamefont {{dos}~{Santos}~{D}ias}\ \emph
  {et~al.}(2015)\citenamefont {{dos}~{Santos}~{D}ias}, \citenamefont
  {Schweflinghaus}, \citenamefont {Bl\"ugel},\ and\ \citenamefont
  {Lounis}}]{dosSantosDias:2015}%
  \BibitemOpen
  \bibfield  {author} {\bibinfo {author} {\bibfnamefont {M.}~\bibnamefont
  {{dos}~{Santos}~{D}ias}}, \bibinfo {author} {\bibfnamefont {B.}~\bibnamefont
  {Schweflinghaus}}, \bibinfo {author} {\bibfnamefont {S.}~\bibnamefont
  {Bl\"ugel}}, \ and\ \bibinfo {author} {\bibfnamefont {S.}~\bibnamefont
  {Lounis}},\ }\href@noop {} {\bibfield  {journal} {\bibinfo  {journal} {Phys.
  Rev. B}\ }\textbf {\bibinfo {volume} {91}},\ \bibinfo {pages} {075405}
  (\bibinfo {year} {2015})}\BibitemShut {NoStop}%
\bibitem [{\citenamefont {Wysocki}\ \emph {et~al.}(2017)\citenamefont
  {Wysocki}, \citenamefont {Valmispild}, \citenamefont {Kutepov}, \citenamefont
  {Sharma}, \citenamefont {Dewhurst}, \citenamefont {Gross}, \citenamefont
  {Lichtenstein},\ and\ \citenamefont {Antropov}}]{Wysocki:2017}%
  \BibitemOpen
  \bibfield  {author} {\bibinfo {author} {\bibfnamefont {A.}~\bibnamefont
  {Wysocki}}, \bibinfo {author} {\bibfnamefont {V.}~\bibnamefont {Valmispild}},
  \bibinfo {author} {\bibfnamefont {A.}~\bibnamefont {Kutepov}}, \bibinfo
  {author} {\bibfnamefont {S.}~\bibnamefont {Sharma}}, \bibinfo {author}
  {\bibfnamefont {J.}~\bibnamefont {Dewhurst}}, \bibinfo {author}
  {\bibfnamefont {E.}~\bibnamefont {Gross}}, \bibinfo {author} {\bibfnamefont
  {A.}~\bibnamefont {Lichtenstein}}, \ and\ \bibinfo {author} {\bibfnamefont
  {V.}~\bibnamefont {Antropov}},\ }\href@noop {} {\bibfield  {journal}
  {\bibinfo  {journal} {Phys. Rev. B}\ }\textbf {\bibinfo {volume} {96}},\
  \bibinfo {pages} {184418} (\bibinfo {year} {2017})}\BibitemShut {NoStop}%
\bibitem [{\citenamefont {Cao}\ \emph {et~al.}(2018)\citenamefont {Cao},
  \citenamefont {Lambert}, \citenamefont {Radaelli},\ and\ \citenamefont
  {Giustino}}]{Cao:2018}%
  \BibitemOpen
  \bibfield  {author} {\bibinfo {author} {\bibfnamefont {K.}~\bibnamefont
  {Cao}}, \bibinfo {author} {\bibfnamefont {H.}~\bibnamefont {Lambert}},
  \bibinfo {author} {\bibfnamefont {P.}~\bibnamefont {Radaelli}}, \ and\
  \bibinfo {author} {\bibfnamefont {F.}~\bibnamefont {Giustino}},\ }\href@noop
  {} {\bibfield  {journal} {\bibinfo  {journal} {Phys. Rev. B}\ }\textbf
  {\bibinfo {volume} {97}},\ \bibinfo {pages} {024420} (\bibinfo {year}
  {2018})}\BibitemShut {NoStop}%
\bibitem [{\citenamefont {Aryasetiawan}\ and\ \citenamefont
  {Karlsson}(1999)}]{Aryasetiawan:1999}%
  \BibitemOpen
  \bibfield  {author} {\bibinfo {author} {\bibfnamefont {F.}~\bibnamefont
  {Aryasetiawan}}\ and\ \bibinfo {author} {\bibfnamefont {K.}~\bibnamefont
  {Karlsson}},\ }\href@noop {} {\bibfield  {journal} {\bibinfo  {journal}
  {Phys. Rev. B}\ }\textbf {\bibinfo {volume} {60}},\ \bibinfo {pages} {7419}
  (\bibinfo {year} {1999})}\BibitemShut {NoStop}%
\bibitem [{\citenamefont {Karlsson}\ and\ \citenamefont
  {Aryasetiawan}(2000)}]{Karlsson:2000}%
  \BibitemOpen
  \bibfield  {author} {\bibinfo {author} {\bibfnamefont {K.}~\bibnamefont
  {Karlsson}}\ and\ \bibinfo {author} {\bibfnamefont {F.}~\bibnamefont
  {Aryasetiawan}},\ }\href@noop {} {\bibfield  {journal} {\bibinfo  {journal}
  {Phys. Rev. B}\ }\textbf {\bibinfo {volume} {62}},\ \bibinfo {pages} {3006}
  (\bibinfo {year} {2000})}\BibitemShut {NoStop}%
\bibitem [{\citenamefont {Kotani}\ and\ \citenamefont {van
  {S}chilfgaarde}(2008)}]{Kotani:2008}%
  \BibitemOpen
  \bibfield  {author} {\bibinfo {author} {\bibfnamefont {T.}~\bibnamefont
  {Kotani}}\ and\ \bibinfo {author} {\bibfnamefont {M.}~\bibnamefont {van
  {S}chilfgaarde}},\ }\href@noop {} {\bibfield  {journal} {\bibinfo  {journal}
  {J. Phys.: Condens. Matter}\ }\textbf {\bibinfo {volume} {20}},\ \bibinfo
  {pages} {295214} (\bibinfo {year} {2008})}\BibitemShut {NoStop}%
\bibitem [{\citenamefont {\c{S}a\c{s}io\u{g}lu}\ \emph
  {et~al.}(2010)\citenamefont {\c{S}a\c{s}io\u{g}lu}, \citenamefont
  {Schindlmayr}, \citenamefont {Friedrich}, \citenamefont {Freimuth},\ and\
  \citenamefont {S.Bl\"ugel}}]{Sasioglu:2010}%
  \BibitemOpen
  \bibfield  {author} {\bibinfo {author} {\bibnamefont {\c{S}a\c{s}io\u{g}lu}},
  \bibinfo {author} {\bibfnamefont {A.}~\bibnamefont {Schindlmayr}}, \bibinfo
  {author} {\bibfnamefont {C.}~\bibnamefont {Friedrich}}, \bibinfo {author}
  {\bibfnamefont {F.}~\bibnamefont {Freimuth}}, \ and\ \bibinfo {author}
  {\bibnamefont {S.Bl\"ugel}},\ }\href@noop {} {\bibfield  {journal} {\bibinfo
  {journal} {Phys. Rev. B}\ }\textbf {\bibinfo {volume} {81}},\ \bibinfo
  {pages} {054434} (\bibinfo {year} {2010})}\BibitemShut {NoStop}%
\bibitem [{\citenamefont {M\"uller}\ \emph {et~al.}(2016)\citenamefont
  {M\"uller}, \citenamefont {Friedrich},\ and\ \citenamefont
  {Bl\"ugel}}]{Muller:2016}%
  \BibitemOpen
  \bibfield  {author} {\bibinfo {author} {\bibfnamefont {M.}~\bibnamefont
  {M\"uller}}, \bibinfo {author} {\bibfnamefont {C.}~\bibnamefont {Friedrich}},
  \ and\ \bibinfo {author} {\bibfnamefont {S.}~\bibnamefont {Bl\"ugel}},\
  }\href@noop {} {\bibfield  {journal} {\bibinfo  {journal} {Phys. Rev. B}\
  }\textbf {\bibinfo {volume} {94}},\ \bibinfo {pages} {064433} (\bibinfo
  {year} {2016})}\BibitemShut {NoStop}%
\bibitem [{\citenamefont {Runge}\ and\ \citenamefont
  {Gross}(1984)}]{Runge:1984}%
  \BibitemOpen
  \bibfield  {author} {\bibinfo {author} {\bibfnamefont {E.}~\bibnamefont
  {Runge}}\ and\ \bibinfo {author} {\bibfnamefont {E.}~\bibnamefont {Gross}},\
  }\href@noop {} {\bibfield  {journal} {\bibinfo  {journal} {Phys. Rev. Lett.}\
  }\textbf {\bibinfo {volume} {52}},\ \bibinfo {pages} {997} (\bibinfo {year}
  {1984})}\BibitemShut {NoStop}%
\bibitem [{\citenamefont {Marques}\ \emph {et~al.}(2012)\citenamefont
  {Marques}, \citenamefont {Maitra}, \citenamefont {Nogueira}, \citenamefont
  {Gross},\ and\ \citenamefont {Rubio}}]{Marques:2012}%
  \BibitemOpen
  \bibinfo {editor} {\bibfnamefont {M.~A.~L.}\ \bibnamefont {Marques}},
  \bibinfo {editor} {\bibfnamefont {N.~T.}\ \bibnamefont {Maitra}}, \bibinfo
  {editor} {\bibfnamefont {F.~M.~S.}\ \bibnamefont {Nogueira}}, \bibinfo
  {editor} {\bibfnamefont {E.~K.~U.}\ \bibnamefont {Gross}}, \ and\ \bibinfo
  {editor} {\bibfnamefont {A.}~\bibnamefont {Rubio}},\ eds.,\ \href@noop {}
  {\emph {\bibinfo {title} {{F}undamentals of {T}ime-{D}ependent {D}ensity
  {F}unctional {T}heory}}},\ Vol.\ \bibinfo {volume} {837}\ (\bibinfo
  {publisher} {{L}ecture {N}otes in {Physics}, {S}pringer-{V}erlag},\ \bibinfo
  {address} {Berlin Heidelbnerg},\ \bibinfo {year} {2012})\BibitemShut
  {NoStop}%
\bibitem [{\citenamefont {Baroni}\ and\ \citenamefont {Gebauer}(
  390)}]{Baroni:2012}%
  \BibitemOpen
  \bibfield  {author} {\bibinfo {author} {\bibfnamefont {S.}~\bibnamefont
  {Baroni}}\ and\ \bibinfo {author} {\bibfnamefont {R.}~\bibnamefont
  {Gebauer}},\ }\href@noop {} {\emph {\bibinfo {title} {The Liouville-Lanczos
  {A}pproach to {T}ime-{D}ependent {D}ensity-{F}unctional ({P}erturbation)
  {Theory}}}}\ (\bibinfo {year} {Ref.~\cite{Marques:2012}, chapter 19,
  p.~375-390})\BibitemShut {NoStop}%
\bibitem [{\citenamefont {Rocca}\ \emph {et~al.}(2008)\citenamefont {Rocca},
  \citenamefont {Gebauer}, \citenamefont {Saad},\ and\ \citenamefont
  {Baroni}}]{Rocca:2008}%
  \BibitemOpen
  \bibfield  {author} {\bibinfo {author} {\bibfnamefont {D.}~\bibnamefont
  {Rocca}}, \bibinfo {author} {\bibfnamefont {R.}~\bibnamefont {Gebauer}},
  \bibinfo {author} {\bibfnamefont {Y.}~\bibnamefont {Saad}}, \ and\ \bibinfo
  {author} {\bibfnamefont {S.}~\bibnamefont {Baroni}},\ }\href@noop {}
  {\bibfield  {journal} {\bibinfo  {journal} {J. Chem. Phys.}\ }\textbf
  {\bibinfo {volume} {128}},\ \bibinfo {pages} {154105} (\bibinfo {year}
  {2008})}\BibitemShut {NoStop}%
\bibitem [{\citenamefont {Timrov}\ \emph {et~al.}(2013)\citenamefont {Timrov},
  \citenamefont {Vast}, \citenamefont {Gebauer},\ and\ \citenamefont
  {Baroni}}]{Timrov:2013}%
  \BibitemOpen
  \bibfield  {author} {\bibinfo {author} {\bibfnamefont {I.}~\bibnamefont
  {Timrov}}, \bibinfo {author} {\bibfnamefont {N.}~\bibnamefont {Vast}},
  \bibinfo {author} {\bibfnamefont {R.}~\bibnamefont {Gebauer}}, \ and\
  \bibinfo {author} {\bibfnamefont {S.}~\bibnamefont {Baroni}},\ }\href@noop {}
  {\bibfield  {journal} {\bibinfo  {journal} {Phys. Rev. B}\ }\textbf {\bibinfo
  {volume} {88}},\ \bibinfo {pages} {064301} (\bibinfo {year} {2013})},\
  \bibinfo {note} {{\it ibid.} {\bf 91}, 139901 (2015).}\BibitemShut {Stop}%
\bibitem [{\citenamefont {Baroni}\ \emph {et~al.}(1987)\citenamefont {Baroni},
  \citenamefont {Giannozzi},\ and\ \citenamefont {Testa}}]{Baroni:1987}%
  \BibitemOpen
  \bibfield  {author} {\bibinfo {author} {\bibfnamefont {S.}~\bibnamefont
  {Baroni}}, \bibinfo {author} {\bibfnamefont {P.}~\bibnamefont {Giannozzi}}, \
  and\ \bibinfo {author} {\bibfnamefont {A.}~\bibnamefont {Testa}},\
  }\href@noop {} {\bibfield  {journal} {\bibinfo  {journal} {Phys. Rev. Lett.}\
  }\textbf {\bibinfo {volume} {58}},\ \bibinfo {pages} {1861} (\bibinfo {year}
  {1987})}\BibitemShut {NoStop}%
\bibitem [{\citenamefont {Baroni}\ \emph {et~al.}(2001)\citenamefont {Baroni},
  \citenamefont {de~Gironcoli}, \citenamefont {Corso},\ and\ \citenamefont
  {Giannozzi}}]{Baroni:2001}%
  \BibitemOpen
  \bibfield  {author} {\bibinfo {author} {\bibfnamefont {S.}~\bibnamefont
  {Baroni}}, \bibinfo {author} {\bibfnamefont {S.}~\bibnamefont
  {de~Gironcoli}}, \bibinfo {author} {\bibfnamefont {A.~D.}\ \bibnamefont
  {Corso}}, \ and\ \bibinfo {author} {\bibfnamefont {P.}~\bibnamefont
  {Giannozzi}},\ }\href@noop {} {\bibfield  {journal} {\bibinfo  {journal}
  {Rev. Mod. Phys.}\ }\textbf {\bibinfo {volume} {73}},\ \bibinfo {pages} {515}
  (\bibinfo {year} {2001})}\BibitemShut {NoStop}%
\bibitem [{\citenamefont {Halpern}\ and\ \citenamefont
  {Johnson}(1938)}]{Halpern:1938aa}%
  \BibitemOpen
  \bibfield  {author} {\bibinfo {author} {\bibfnamefont {O.}~\bibnamefont
  {Halpern}}\ and\ \bibinfo {author} {\bibfnamefont {M.}~\bibnamefont
  {Johnson}},\ }\href@noop {} {\bibfield  {journal} {\bibinfo  {journal} {Phys.
  Rev.}\ }\textbf {\bibinfo {volume} {55}},\ \bibinfo {pages} {898} (\bibinfo
  {year} {1938})}\BibitemShut {NoStop}%
\bibitem [{\citenamefont {Blume}(1963)}]{Blume:1963aa}%
  \BibitemOpen
  \bibfield  {author} {\bibinfo {author} {\bibfnamefont {M.}~\bibnamefont
  {Blume}},\ }\href@noop {} {\bibfield  {journal} {\bibinfo  {journal} {Phys.
  Rev.}\ }\textbf {\bibinfo {volume} {130}},\ \bibinfo {pages} {1670} (\bibinfo
  {year} {1963})}\BibitemShut {NoStop}%
\bibitem [{Not({\natexlab{a}})}]{Note:notations_circ}%
  \BibitemOpen
  \href@noop {} {}  \bibinfo {note} {{H}ereafter with the
  superscript ``$\!\!\!\phantom{a}^\circ$'' we denote quantities which refer to
  the ground state of the system.}\BibitemShut {Stop}%
\bibitem [{\citenamefont {Gorni}(3342)}]{Gorni:2016}%
  \BibitemOpen
  \bibfield  {author} {\bibinfo {author} {\bibfnamefont {T.}~\bibnamefont
  {Gorni}},\ }\emph {\bibinfo {title} {{S}pin-fluctuation spectra in magnetic
  systems: a novel approach based on TDDFT}},\ \href
  {http://hdl.handle.net/20.500.11767/43342} {Ph.D. thesis},\ \bibinfo
  {school} {Scuola Internazionale Superiore di Studi Avanzati (SISSA), Trieste,
  Italy} (\bibinfo {year} {2016,
  http://hdl.handle.net/20.500.11767/43342})\BibitemShut {NoStop}%
\bibitem [{Not({\natexlab{b}})}]{Note:notation_Vxc}%
  \BibitemOpen
  \href@noop {} {} \bibinfo {note} {{W}e note that in the
  second term of Eq.~\eqref{eq:v_XC_q} we symbolically mean a scalar product,
  while in the second term of Eq.~\eqref{eq:b_XC_q} we symbolically mean a
  matrix-vector multiplication.}\BibitemShut {Stop}%
\bibitem [{Not({\natexlab{c}})}]{Note:density_property}%
  \BibitemOpen
  \href@noop {} {} \bibinfo {note} {{T}hese relations are a
  consequence of the fact that charge- and magnetization-density responses are
  real functions in space and time.}\BibitemShut {Stop}%
\bibitem [{Not({\natexlab{d}})}]{Note:ext_pot}%
  \BibitemOpen
  \href@noop {} {} \bibinfo {note} {{S}ee, {\it e.g.}, the
  analogue in the case of the absorption spectroscopy, Eq.~(16) in
  Ref.~\cite{Rocca:2008}.}\BibitemShut {Stop}%
\bibitem [{\citenamefont {Gokhale}\ \emph {et~al.}(1992)\citenamefont
  {Gokhale}, \citenamefont {Ormeci},\ and\ \citenamefont
  {Mills}}]{Gokhale:1992}%
  \BibitemOpen
  \bibfield  {author} {\bibinfo {author} {\bibfnamefont {M.}~\bibnamefont
  {Gokhale}}, \bibinfo {author} {\bibfnamefont {A.}~\bibnamefont {Ormeci}}, \
  and\ \bibinfo {author} {\bibfnamefont {D.}~\bibnamefont {Mills}},\
  }\href@noop {} {\bibfield  {journal} {\bibinfo  {journal} {Phys. Rev. B}\
  }\textbf {\bibinfo {volume} {46}},\ \bibinfo {pages} {8978} (\bibinfo {year}
  {1992})}\BibitemShut {NoStop}%
\bibitem [{Not({\natexlab{e}})}]{Note:SOC}%
  \BibitemOpen
  \href@noop {} {} \bibinfo {note} {{T}he inclusion of
  spin-orbit coupling (SOC) in the Liouville-Lanczos approach when
  non-collinear magnetism is explicitly contemplated is trivial, because SOC is
  time-reversal invariant~\cite{Gorni:2016}. In practice this means that the
  application of the time-reversal operator $\hat{\mathrm{T}}$ to the SOC
  potential on the left-hand side of
  Eq.~\eqref{eq:KS_eq_antiresonant_projected_q} [and in
  Eq.~\eqref{eq:KS_eq_resonant_projected_q}] will not change it. More details
  about the extension of our formalism to SOC will be given
  elsewhere.}\BibitemShut {Stop}%
\bibitem [{\citenamefont {Malcioi\u{g}lu}\ \emph {et~al.}(2011)\citenamefont
  {Malcioi\u{g}lu}, \citenamefont {Gebauer}, \citenamefont {Rocca},\ and\
  \citenamefont {Baroni}}]{Malcioglu:2011}%
  \BibitemOpen
  \bibfield  {author} {\bibinfo {author} {\bibfnamefont {O.}~\bibnamefont
  {Malcioi\u{g}lu}}, \bibinfo {author} {\bibfnamefont {R.}~\bibnamefont
  {Gebauer}}, \bibinfo {author} {\bibfnamefont {D.}~\bibnamefont {Rocca}}, \
  and\ \bibinfo {author} {\bibfnamefont {S.}~\bibnamefont {Baroni}},\
  }\href@noop {} {\bibfield  {journal} {\bibinfo  {journal} {Comput. Phys.
  Commun.}\ }\textbf {\bibinfo {volume} {182}},\ \bibinfo {pages} {1744}
  (\bibinfo {year} {2011})}\BibitemShut {NoStop}%
\bibitem [{\citenamefont {Ge}\ \emph {et~al.}(2014)\citenamefont {Ge},
  \citenamefont {Binnie}, \citenamefont {Rocca}, \citenamefont {Gebauer},\ and\
  \citenamefont {Baroni}}]{Ge:2014}%
  \BibitemOpen
  \bibfield  {author} {\bibinfo {author} {\bibfnamefont {X.}~\bibnamefont
  {Ge}}, \bibinfo {author} {\bibfnamefont {S.~J.}\ \bibnamefont {Binnie}},
  \bibinfo {author} {\bibfnamefont {D.}~\bibnamefont {Rocca}}, \bibinfo
  {author} {\bibfnamefont {R.}~\bibnamefont {Gebauer}}, \ and\ \bibinfo
  {author} {\bibfnamefont {S.}~\bibnamefont {Baroni}},\ }\href@noop {}
  {\bibfield  {journal} {\bibinfo  {journal} {Comput. Phys. Commun.}\ }\textbf
  {\bibinfo {volume} {185}},\ \bibinfo {pages} {2080} (\bibinfo {year}
  {2014})}\BibitemShut {NoStop}%
\bibitem [{\citenamefont {Saad}(2003)}]{Saad:2003}%
  \BibitemOpen
  \bibfield  {author} {\bibinfo {author} {\bibfnamefont {Y.}~\bibnamefont
  {Saad}},\ }\href@noop {} {\emph {\bibinfo {title} {Iterative Methods for
  Sparse Linear Systems}}},\ \bibinfo {edition} {2nd}\ ed.\ (\bibinfo
  {publisher} {SIAM},\ \bibinfo {address} {Philadelphia},\ \bibinfo {year}
  {2003})\BibitemShut {NoStop}%
\bibitem [{\citenamefont {Gr\"uning}\ \emph {et~al.}(2011)\citenamefont
  {Gr\"uning}, \citenamefont {Marini},\ and\ \citenamefont
  {Gonze}}]{Gruning:2011}%
  \BibitemOpen
  \bibfield  {author} {\bibinfo {author} {\bibfnamefont {M.}~\bibnamefont
  {Gr\"uning}}, \bibinfo {author} {\bibfnamefont {A.}~\bibnamefont {Marini}}, \
  and\ \bibinfo {author} {\bibfnamefont {X.}~\bibnamefont {Gonze}},\
  }\href@noop {} {\bibfield  {journal} {\bibinfo  {journal} {Comput. Math.
  Sci.}\ }\textbf {\bibinfo {volume} {50}},\ \bibinfo {pages} {2148} (\bibinfo
  {year} {2011})}\BibitemShut {NoStop}%
\bibitem [{\citenamefont {Giannozzi}\ \emph {et~al.}(2009)\citenamefont
  {Giannozzi}, \citenamefont {Baroni}, \citenamefont {Bonini}, \citenamefont
  {Calandra}, \citenamefont {Car}, \citenamefont {Cavazzoni}, \citenamefont
  {Ceresoli}, \citenamefont {Chiarotti}, \citenamefont {Cococcioni},
  \citenamefont {Dabo}, \citenamefont {Dal~Corso}, \citenamefont
  {De~Gironcoli}, \citenamefont {Fabris}, \citenamefont {Fratesi},
  \citenamefont {Gebauer}, \citenamefont {Gerstmann}, \citenamefont
  {Gougoussis}, \citenamefont {Kokalj}, \citenamefont {Lazzeri}, \citenamefont
  {Martin-Samos}, \citenamefont {Marzari}, \citenamefont {Mauri}, \citenamefont
  {Mazzarello}, \citenamefont {Paolini}, \citenamefont {Pasquarello},
  \citenamefont {Paulatto}, \citenamefont {Sbraccia}, \citenamefont {Scandolo},
  \citenamefont {Sclauzero}, \citenamefont {Seitsonen}, \citenamefont
  {Smogunov}, \citenamefont {Umari},\ and\ \citenamefont
  {Wentzcovitch}}]{Giannozzi:2009}%
  \BibitemOpen
  \bibfield  {author} {\bibinfo {author} {\bibfnamefont {P.}~\bibnamefont
  {Giannozzi}}, \bibinfo {author} {\bibfnamefont {S.}~\bibnamefont {Baroni}},
  \bibinfo {author} {\bibfnamefont {N.}~\bibnamefont {Bonini}}, \bibinfo
  {author} {\bibfnamefont {M.}~\bibnamefont {Calandra}}, \bibinfo {author}
  {\bibfnamefont {R.}~\bibnamefont {Car}}, \bibinfo {author} {\bibfnamefont
  {C.}~\bibnamefont {Cavazzoni}}, \bibinfo {author} {\bibfnamefont
  {D.}~\bibnamefont {Ceresoli}}, \bibinfo {author} {\bibfnamefont
  {G.}~\bibnamefont {Chiarotti}}, \bibinfo {author} {\bibfnamefont
  {M.}~\bibnamefont {Cococcioni}}, \bibinfo {author} {\bibfnamefont
  {I.}~\bibnamefont {Dabo}}, \bibinfo {author} {\bibfnamefont {A.}~\bibnamefont
  {Dal~Corso}}, \bibinfo {author} {\bibfnamefont {S.}~\bibnamefont
  {De~Gironcoli}}, \bibinfo {author} {\bibfnamefont {S.}~\bibnamefont
  {Fabris}}, \bibinfo {author} {\bibfnamefont {G.}~\bibnamefont {Fratesi}},
  \bibinfo {author} {\bibfnamefont {R.}~\bibnamefont {Gebauer}}, \bibinfo
  {author} {\bibfnamefont {U.}~\bibnamefont {Gerstmann}}, \bibinfo {author}
  {\bibfnamefont {C.}~\bibnamefont {Gougoussis}}, \bibinfo {author}
  {\bibfnamefont {A.}~\bibnamefont {Kokalj}}, \bibinfo {author} {\bibfnamefont
  {M.}~\bibnamefont {Lazzeri}}, \bibinfo {author} {\bibfnamefont
  {L.}~\bibnamefont {Martin-Samos}}, \bibinfo {author} {\bibfnamefont
  {N.}~\bibnamefont {Marzari}}, \bibinfo {author} {\bibfnamefont
  {F.}~\bibnamefont {Mauri}}, \bibinfo {author} {\bibfnamefont
  {R.}~\bibnamefont {Mazzarello}}, \bibinfo {author} {\bibfnamefont
  {S.}~\bibnamefont {Paolini}}, \bibinfo {author} {\bibfnamefont
  {A.}~\bibnamefont {Pasquarello}}, \bibinfo {author} {\bibfnamefont
  {L.}~\bibnamefont {Paulatto}}, \bibinfo {author} {\bibfnamefont
  {C.}~\bibnamefont {Sbraccia}}, \bibinfo {author} {\bibfnamefont
  {S.}~\bibnamefont {Scandolo}}, \bibinfo {author} {\bibfnamefont
  {G.}~\bibnamefont {Sclauzero}}, \bibinfo {author} {\bibfnamefont
  {A.}~\bibnamefont {Seitsonen}}, \bibinfo {author} {\bibfnamefont
  {A.}~\bibnamefont {Smogunov}}, \bibinfo {author} {\bibfnamefont
  {P.}~\bibnamefont {Umari}}, \ and\ \bibinfo {author} {\bibfnamefont
  {R.}~\bibnamefont {Wentzcovitch}},\ }\href {http://www.quantum-espresso.org}
  {\bibfield  {journal} {\bibinfo  {journal} {J. Phys.: Condens. Matter}\
  }\textbf {\bibinfo {volume} {21}},\ \bibinfo {pages} {395502} (\bibinfo
  {year} {2009})}\BibitemShut {NoStop}%
\bibitem [{\citenamefont {Giannozzi}\ \emph {et~al.}(2017)\citenamefont
  {Giannozzi}, \citenamefont {Andreussi}, \citenamefont {Brumme}, \citenamefont
  {Bunau}, \citenamefont {Buongiorno~Nardelli}, \citenamefont {Calandra},
  \citenamefont {Car}, \citenamefont {Cavazzoni}, \citenamefont {Ceresoli},
  \citenamefont {Cococcioni}, \citenamefont {Colonna}, \citenamefont
  {Carnimeo}, \citenamefont {Dal~Corso}, \citenamefont {de~Gironcoli},
  \citenamefont {Delugas}, \citenamefont {Di{S}tasio~{J}r.}, \citenamefont
  {Ferretti}, \citenamefont {Floris}, \citenamefont {Fratesi}, \citenamefont
  {Fugallo}, \citenamefont {Gebauer}, \citenamefont {Gerstmann}, \citenamefont
  {Giustino}, \citenamefont {Gorni}, \citenamefont {Jia}, \citenamefont
  {Kawamura}, \citenamefont {Ko}, \citenamefont {Kokalj}, \citenamefont
  {K\"{u}\c{c}\"{u}kbenli}, \citenamefont {Lazzeri}, \citenamefont {Marsili},
  \citenamefont {Marzari}, \citenamefont {Mauri}, \citenamefont {Nguyen},
  \citenamefont {Nguyen}, \citenamefont {Otero-de-la {R}osa}, \citenamefont
  {Paulatto}, \citenamefont {Ponc\'e}, \citenamefont {Rocca}, \citenamefont
  {Sabatini}, \citenamefont {Santra}, \citenamefont {Schlipf}, \citenamefont
  {Seitsonen}, \citenamefont {Smogunov}, \citenamefont {Timrov}, \citenamefont
  {Thonhauser}, \citenamefont {Umari}, \citenamefont {Vast},\ and\
  \citenamefont {Baroni}}]{Giannozzi:2017}%
  \BibitemOpen
  \bibfield  {author} {\bibinfo {author} {\bibfnamefont {P.}~\bibnamefont
  {Giannozzi}}, \bibinfo {author} {\bibfnamefont {O.}~\bibnamefont
  {Andreussi}}, \bibinfo {author} {\bibfnamefont {T.}~\bibnamefont {Brumme}},
  \bibinfo {author} {\bibfnamefont {O.}~\bibnamefont {Bunau}}, \bibinfo
  {author} {\bibfnamefont {M.}~\bibnamefont {Buongiorno~Nardelli}}, \bibinfo
  {author} {\bibfnamefont {M.}~\bibnamefont {Calandra}}, \bibinfo {author}
  {\bibfnamefont {R.}~\bibnamefont {Car}}, \bibinfo {author} {\bibfnamefont
  {C.}~\bibnamefont {Cavazzoni}}, \bibinfo {author} {\bibfnamefont
  {D.}~\bibnamefont {Ceresoli}}, \bibinfo {author} {\bibfnamefont
  {M.}~\bibnamefont {Cococcioni}}, \bibinfo {author} {\bibfnamefont
  {N.}~\bibnamefont {Colonna}}, \bibinfo {author} {\bibfnamefont
  {I.}~\bibnamefont {Carnimeo}}, \bibinfo {author} {\bibfnamefont
  {A.}~\bibnamefont {Dal~Corso}}, \bibinfo {author} {\bibfnamefont
  {S.}~\bibnamefont {de~Gironcoli}}, \bibinfo {author} {\bibfnamefont
  {P.}~\bibnamefont {Delugas}}, \bibinfo {author} {\bibfnamefont {R.~A.}\
  \bibnamefont {Di{S}tasio~{J}r.}}, \bibinfo {author} {\bibfnamefont
  {A.}~\bibnamefont {Ferretti}}, \bibinfo {author} {\bibfnamefont
  {A.}~\bibnamefont {Floris}}, \bibinfo {author} {\bibfnamefont
  {G.}~\bibnamefont {Fratesi}}, \bibinfo {author} {\bibfnamefont
  {G.}~\bibnamefont {Fugallo}}, \bibinfo {author} {\bibfnamefont
  {R.}~\bibnamefont {Gebauer}}, \bibinfo {author} {\bibfnamefont
  {U.}~\bibnamefont {Gerstmann}}, \bibinfo {author} {\bibfnamefont
  {F.}~\bibnamefont {Giustino}}, \bibinfo {author} {\bibfnamefont
  {T.}~\bibnamefont {Gorni}}, \bibinfo {author} {\bibfnamefont
  {J.}~\bibnamefont {Jia}}, \bibinfo {author} {\bibfnamefont {M.}~\bibnamefont
  {Kawamura}}, \bibinfo {author} {\bibfnamefont {H.-Y.}\ \bibnamefont {Ko}},
  \bibinfo {author} {\bibfnamefont {A.}~\bibnamefont {Kokalj}}, \bibinfo
  {author} {\bibfnamefont {E.}~\bibnamefont {K\"{u}\c{c}\"{u}kbenli}}, \bibinfo
  {author} {\bibfnamefont {M.}~\bibnamefont {Lazzeri}}, \bibinfo {author}
  {\bibfnamefont {M.}~\bibnamefont {Marsili}}, \bibinfo {author} {\bibfnamefont
  {N.}~\bibnamefont {Marzari}}, \bibinfo {author} {\bibfnamefont
  {F.}~\bibnamefont {Mauri}}, \bibinfo {author} {\bibfnamefont {N.~L.}\
  \bibnamefont {Nguyen}}, \bibinfo {author} {\bibfnamefont {H.-V.}\
  \bibnamefont {Nguyen}}, \bibinfo {author} {\bibfnamefont {A.}~\bibnamefont
  {Otero-de-la {R}osa}}, \bibinfo {author} {\bibfnamefont {L.}~\bibnamefont
  {Paulatto}}, \bibinfo {author} {\bibfnamefont {S.}~\bibnamefont {Ponc\'e}},
  \bibinfo {author} {\bibfnamefont {D.}~\bibnamefont {Rocca}}, \bibinfo
  {author} {\bibfnamefont {R.}~\bibnamefont {Sabatini}}, \bibinfo {author}
  {\bibfnamefont {B.}~\bibnamefont {Santra}}, \bibinfo {author} {\bibfnamefont
  {M.}~\bibnamefont {Schlipf}}, \bibinfo {author} {\bibfnamefont
  {A.}~\bibnamefont {Seitsonen}}, \bibinfo {author} {\bibfnamefont
  {A.}~\bibnamefont {Smogunov}}, \bibinfo {author} {\bibfnamefont
  {I.}~\bibnamefont {Timrov}}, \bibinfo {author} {\bibfnamefont
  {T.}~\bibnamefont {Thonhauser}}, \bibinfo {author} {\bibfnamefont
  {P.}~\bibnamefont {Umari}}, \bibinfo {author} {\bibfnamefont
  {N.}~\bibnamefont {Vast}}, \ and\ \bibinfo {author} {\bibfnamefont
  {S.}~\bibnamefont {Baroni}},\ }\href@noop {} {\bibfield  {journal} {\bibinfo
  {journal} {J. Phys.: Condens. Matter}\ }\textbf {\bibinfo {volume} {29}},\
  \bibinfo {pages} {465901} (\bibinfo {year} {2017})}\BibitemShut {NoStop}%
\bibitem [{\citenamefont {van Setten}\ \emph {et~al.}(2018)\citenamefont {van
  Setten}, \citenamefont {Giantomassi}, \citenamefont {Bousquet}, \citenamefont
  {Verstraete}, \citenamefont {Hamann}, \citenamefont {Gonze},\ and\
  \citenamefont {Rignanese}}]{Setten:2018}%
  \BibitemOpen
  \bibfield  {author} {\bibinfo {author} {\bibfnamefont {M.}~\bibnamefont {van
  Setten}}, \bibinfo {author} {\bibfnamefont {M.}~\bibnamefont {Giantomassi}},
  \bibinfo {author} {\bibfnamefont {E.}~\bibnamefont {Bousquet}}, \bibinfo
  {author} {\bibfnamefont {M.}~\bibnamefont {Verstraete}}, \bibinfo {author}
  {\bibfnamefont {D.}~\bibnamefont {Hamann}}, \bibinfo {author} {\bibfnamefont
  {X.}~\bibnamefont {Gonze}}, \ and\ \bibinfo {author} {\bibfnamefont {G.-M.}\
  \bibnamefont {Rignanese}},\ }\href {www.pseudo-dojo.org/} {\bibfield
  {journal} {\bibinfo  {journal} {Comput. Phys. Commun.}\ }\textbf {\bibinfo
  {volume} {226}},\ \bibinfo {pages} {39} (\bibinfo {year} {2018})}\BibitemShut
  {NoStop}%
\bibitem [{Doj()}]{DojoPP:2018}%
  \BibitemOpen
  \href@noop {} {}\bibinfo {note} {We have used the following pseudopotentials:
  \texttt{Fe.upf} and \texttt{Ni.upf}, both of the ONCVPSP v0.3 type,
  www.pseudo-dojo.org}\BibitemShut {NoStop}%
\bibitem [{\citenamefont {Basinski}\ \emph {et~al.}(1955)\citenamefont
  {Basinski}, \citenamefont {Hume-Rothery},\ and\ \citenamefont
  {Sutton}}]{Basinski:1955}%
  \BibitemOpen
  \bibfield  {author} {\bibinfo {author} {\bibfnamefont {Z.}~\bibnamefont
  {Basinski}}, \bibinfo {author} {\bibfnamefont {W.}~\bibnamefont
  {Hume-Rothery}}, \ and\ \bibinfo {author} {\bibfnamefont {A.}~\bibnamefont
  {Sutton}},\ }\href@noop {} {\bibfield  {journal} {\bibinfo  {journal} {Proc.
  Roy. Soc. A}\ }\textbf {\bibinfo {volume} {229}},\ \bibinfo {pages} {459}
  (\bibinfo {year} {1955})}\BibitemShut {NoStop}%
\bibitem [{\citenamefont {Bandyopadhyaya}\ and\ \citenamefont
  {Gupta}(1977)}]{Bandyopadhyaya:1977}%
  \BibitemOpen
  \bibfield  {author} {\bibinfo {author} {\bibfnamefont {J.}~\bibnamefont
  {Bandyopadhyaya}}\ and\ \bibinfo {author} {\bibfnamefont {K.}~\bibnamefont
  {Gupta}},\ }\href@noop {} {\bibfield  {journal} {\bibinfo  {journal}
  {Cryogenics}\ }\textbf {\bibinfo {volume} {17}},\ \bibinfo {pages} {345}
  (\bibinfo {year} {1977})}\BibitemShut {NoStop}%
\bibitem [{\citenamefont {Crangle}\ and\ \citenamefont
  {Goodman}(1971)}]{Crangle:1971}%
  \BibitemOpen
  \bibfield  {author} {\bibinfo {author} {\bibfnamefont {J.}~\bibnamefont
  {Crangle}}\ and\ \bibinfo {author} {\bibfnamefont {G.}~\bibnamefont
  {Goodman}},\ }\href@noop {} {\bibfield  {journal} {\bibinfo  {journal} {Proc.
  Roy. Soc. A}\ }\textbf {\bibinfo {volume} {321}},\ \bibinfo {pages} {477}
  (\bibinfo {year} {1971})}\BibitemShut {NoStop}%
\bibitem [{\citenamefont {Iota}\ \emph {et~al.}(2007)\citenamefont {Iota},
  \citenamefont {Klepeis}, \citenamefont {Yoo}, \citenamefont {Lang},
  \citenamefont {Haskel},\ and\ \citenamefont {Srajer}}]{Iota:2007}%
  \BibitemOpen
  \bibfield  {author} {\bibinfo {author} {\bibfnamefont {V.}~\bibnamefont
  {Iota}}, \bibinfo {author} {\bibfnamefont {J.-H.}\ \bibnamefont {Klepeis}},
  \bibinfo {author} {\bibfnamefont {C.-S.}\ \bibnamefont {Yoo}}, \bibinfo
  {author} {\bibfnamefont {J.}~\bibnamefont {Lang}}, \bibinfo {author}
  {\bibfnamefont {D.}~\bibnamefont {Haskel}}, \ and\ \bibinfo {author}
  {\bibfnamefont {G.}~\bibnamefont {Srajer}},\ }\href@noop {} {\bibfield
  {journal} {\bibinfo  {journal} {Appl. Phys. Let.}\ }\textbf {\bibinfo
  {volume} {90}},\ \bibinfo {pages} {042505} (\bibinfo {year}
  {2007})}\BibitemShut {NoStop}%
\bibitem [{\citenamefont {Motornyi}\ \emph {et~al.}(pear)\citenamefont
  {Motornyi}, \citenamefont {Raynaud}, \citenamefont {Dal~Corso},\ and\
  \citenamefont {Vast}}]{Motornyi:2018}%
  \BibitemOpen
  \bibfield  {author} {\bibinfo {author} {\bibfnamefont {O.}~\bibnamefont
  {Motornyi}}, \bibinfo {author} {\bibfnamefont {M.}~\bibnamefont {Raynaud}},
  \bibinfo {author} {\bibfnamefont {A.}~\bibnamefont {Dal~Corso}}, \ and\
  \bibinfo {author} {\bibfnamefont {N.}~\bibnamefont {Vast}},\ }\href@noop {}
  {\bibfield  {journal} {\bibinfo  {journal} {J. Phys. Conf. Ser.}\ } (\bibinfo
  {year} {2018, to appear})}\BibitemShut {NoStop}%
\bibitem [{\citenamefont {Loong}\ \emph {et~al.}(1984)\citenamefont {Loong},
  \citenamefont {Carpenter}, \citenamefont {Lynn}, \citenamefont {Robinson},\
  and\ \citenamefont {Mook}}]{Loong:1984}%
  \BibitemOpen
  \bibfield  {author} {\bibinfo {author} {\bibfnamefont {C.}~\bibnamefont
  {Loong}}, \bibinfo {author} {\bibfnamefont {J.}~\bibnamefont {Carpenter}},
  \bibinfo {author} {\bibfnamefont {J.}~\bibnamefont {Lynn}}, \bibinfo {author}
  {\bibfnamefont {R.}~\bibnamefont {Robinson}}, \ and\ \bibinfo {author}
  {\bibfnamefont {H.}~\bibnamefont {Mook}},\ }\href@noop {} {\bibfield
  {journal} {\bibinfo  {journal} {J. Appl. Phys.}\ }\textbf {\bibinfo {volume}
  {55}},\ \bibinfo {pages} {1895} (\bibinfo {year} {1984})}\BibitemShut
  {NoStop}%
\bibitem [{\citenamefont {Mook}\ and\ \citenamefont {Paul}(1985)}]{Mook:1985}%
  \BibitemOpen
  \bibfield  {author} {\bibinfo {author} {\bibfnamefont {H.}~\bibnamefont
  {Mook}}\ and\ \bibinfo {author} {\bibfnamefont {D.~M.}\ \bibnamefont
  {Paul}},\ }\href@noop {} {\bibfield  {journal} {\bibinfo  {journal} {Phys.
  Rev. Lett.}\ }\textbf {\bibinfo {volume} {54}},\ \bibinfo {pages} {227}
  (\bibinfo {year} {1985})}\BibitemShut {NoStop}%
\bibitem [{\citenamefont {Eich}\ \emph {et~al.}(2018)\citenamefont {Eich},
  \citenamefont {Pittalis},\ and\ \citenamefont {Vignale}}]{Eich:2018}%
  \BibitemOpen
  \bibfield  {author} {\bibinfo {author} {\bibfnamefont {F.}~\bibnamefont
  {Eich}}, \bibinfo {author} {\bibfnamefont {S.}~\bibnamefont {Pittalis}}, \
  and\ \bibinfo {author} {\bibfnamefont {G.}~\bibnamefont {Vignale}},\
  }\href@noop {} {\bibfield  {journal} {\bibinfo  {journal} {Eur. Phys. J. B}\
  }\textbf {\bibinfo {volume} {91}},\ \bibinfo {pages} {173} (\bibinfo {year}
  {2018})}\BibitemShut {NoStop}%
\bibitem [{\citenamefont {Fu}\ and\ \citenamefont {Ho}(1983)}]{Fu:1983}%
  \BibitemOpen
  \bibfield  {author} {\bibinfo {author} {\bibfnamefont {C.-L.}\ \bibnamefont
  {Fu}}\ and\ \bibinfo {author} {\bibfnamefont {K.-M.}\ \bibnamefont {Ho}},\
  }\href {\doibase 10.1103/PhysRevB.28.5480} {\bibfield  {journal} {\bibinfo
  {journal} {Phys. Rev. B}\ }\textbf {\bibinfo {volume} {28}},\ \bibinfo
  {pages} {5480} (\bibinfo {year} {1983})}\BibitemShut {NoStop}%
\bibitem [{\citenamefont {Methfessel}\ and\ \citenamefont
  {Paxton}(1989)}]{Methfessel:1989}%
  \BibitemOpen
  \bibfield  {author} {\bibinfo {author} {\bibfnamefont {M.}~\bibnamefont
  {Methfessel}}\ and\ \bibinfo {author} {\bibfnamefont {A.}~\bibnamefont
  {Paxton}},\ }\href@noop {} {\bibfield  {journal} {\bibinfo  {journal} {Phys.
  Rev. B}\ }\textbf {\bibinfo {volume} {40}},\ \bibinfo {pages} {3616}
  (\bibinfo {year} {1989})}\BibitemShut {NoStop}%
\bibitem [{\citenamefont {Marzari}\ \emph {et~al.}(1999)\citenamefont
  {Marzari}, \citenamefont {Vanderbilt}, \citenamefont {Vita},\ and\
  \citenamefont {Payne}}]{Marzari:1999}%
  \BibitemOpen
  \bibfield  {author} {\bibinfo {author} {\bibfnamefont {N.}~\bibnamefont
  {Marzari}}, \bibinfo {author} {\bibfnamefont {D.}~\bibnamefont {Vanderbilt}},
  \bibinfo {author} {\bibfnamefont {A.~D.}\ \bibnamefont {Vita}}, \ and\
  \bibinfo {author} {\bibfnamefont {M.~C.}\ \bibnamefont {Payne}},\ }\href@noop
  {} {\bibfield  {journal} {\bibinfo  {journal} {Phys. Rev. Lett.}\ }\textbf
  {\bibinfo {volume} {82}},\ \bibinfo {pages} {3296} (\bibinfo {year}
  {1999})}\BibitemShut {NoStop}%
\bibitem [{\citenamefont {de~Gironcoli}(1995)}]{deGironcoli:1995}%
  \BibitemOpen
  \bibfield  {author} {\bibinfo {author} {\bibfnamefont {S.}~\bibnamefont
  {de~Gironcoli}},\ }\href@noop {} {\bibfield  {journal} {\bibinfo  {journal}
  {Phys. Rev. B}\ }\textbf {\bibinfo {volume} {51}},\ \bibinfo {pages} {6773}
  (\bibinfo {year} {1995})}\BibitemShut {NoStop}%
\bibitem [{Not({\natexlab{f}})}]{Note:notations_theta}%
  \BibitemOpen
  \href@noop {} {} \bibinfo {note} {{T}he $\theta$ function
  can be chosen to be any smooth function which approximates a step-like
  function, because it comes from the condition $\theta(\varepsilon) +
  \theta(-\varepsilon) = 1$ which is used in the derivation of the equations
  \cite{deGironcoli:1995}. For example, in Refs.~\cite{deGironcoli:1995,
  Baroni:2001} it is considered to be equal to $\tilde{\theta}$ which depends
  on the smearing technique which is used. In this work $\theta$ is taken to be
  a rescaled complementary error function, as implemented in the latest
  versions of the \QE package, which makes the numerical stability of the
  calculations more robust.}\BibitemShut {Stop}%
\end{thebibliography}%

%

\end{document}